\documentclass[a4paper,fleqn,usenatbib]{mnras}

\usepackage[T1]{fontenc}
\usepackage{ae,aecompl}

\usepackage{graphicx}
\usepackage{amsmath}
\usepackage{amssymb}

\title[A testbed for cosmological zoom-in simulations]{An observational testbed for cosmological zoom-in simulations: constraining stellar migration in the solar cylinder using asteroseismology}

\author[Verma et al.]{Kuldeep Verma,$^{1}$\thanks{E-mail: kuldeep@phys.au.dk (KV)}
Robert J. J. Grand,$^{2}$\thanks{E-mail: grand@mpa-garching.mpg.de (RG)}
V\'{i}ctor Silva Aguirre,$^{1}$\thanks{E-mail: victor@phys.au.dk (VSA)}
Amalie Stokholm$^{1}$\thanks{E-mail: stokholm@phys.au.dk (AS)}
\\
$^{1}$Stellar Astrophysics Centre, Department of Physics and Astronomy, Aarhus University, Ny Munkegade 120, DK-8000 Aarhus C, Denmark
\\
$^{2}$Max-Planck-Institut f\"{u}r Astrophysik, Karl-Schwarzschild-Str 1, D-85748 Garching, Germany 
}

\date{Accepted XXX. Received YYY; in original form ZZZ}
\pubyear{2021}

\begin{document}
\label{firstpage}
\pagerange{\pageref{firstpage}--\pageref{lastpage}}
\maketitle

\begin{abstract}
Large-scale stellar surveys coupled with recent developments in magneto-hydrodynamical simulations of the formation of Milky Way-mass galaxies provide an unparalleled opportunity to unveil the physical processes driving the evolution of the Galaxy. We developed a framework to compare a variety of parameters with their corresponding predictions from simulations in an unbiased manner, taking into account the selection function of a stellar survey. We applied this framework to a sample of over 7000 stars with asteroseismic, spectroscopic, and astrometric data available, together with 6 simulations from the Auriga project. We found that some simulations are able to produce abundance dichotomies in the $[{\rm Fe}/{\rm H}]-[\alpha/{\rm Fe}]$ plane which look qualitatively similar to observations. The peak of their velocity distributions match the observed data reasonably well, however they predict hotter kinematics in terms of the tails of the distributions and the vertical velocity dispersion. Assuming our simulation sample is representative of Milky Way-like galaxies, we put upper limits of 2.21 and 3.70 kpc on radial migration for young ($< 4$ Gyr) and old ($\in [4, 8]$ Gyr) stellar populations in the solar cylinder. Comparison between the observed and simulated metallicity dispersion as a function of age further constrains migration to about 1.97 and 2.91 kpc for the young and old populations. These results demonstrate the power of our technique to compare numerical simulations with high-dimensional datasets, and paves the way for using the wider field TESS asteroseismic data together with the future generations of simulations to constrain the subgrid models for turbulence, star formation and feedback processes.
\end{abstract}


\section{Introduction}
\label{intro}
We are currently in an unprecedented position to understand one of the most important problems of modern astrophysics, viz. the formation and 
evolution of our host galaxy, the Milky Way. On the one hand, we have rich observational stellar data for a significant fraction of the 
whole sky from various ground- and space-based instruments; for instance spectroscopic data from the RAVE, LAMOST, {\it Gaia}-ESO, GALAH and APOGEE 
surveys \citep[see e.g.][]{rave06,lamo12,geso12,gala15,apog17}, astrometric data from the {\it Gaia} mission \citep[see e.g.][]{gaia16,gaia18} and 
asteroseismic data from the {\it Kepler}/K2 and TESS satellites \citep[see e.g.][]{gill10b,howe14,rick14}. Specifically, asteroseismology enables
us to measure stellar ages to precision better than 20\% \citep[see e.g.][]{silv20}. These data contain complex signatures of the key events 
taking place in the assembly history of the Galaxy. On the other hand, we now have cosmological magneto-hydrodynamical zoom-in simulations of the 
formation of Milky Way-mass galaxies \citep[see e.g.][]{gran17}, which help us in accurately interpreting the observations and provide important 
insights into the physical processes which played critical role in the evolutionary history of the Milky Way.

The early stellar counts in the solar-neighbourhood indicated the presence of two distinct groups of stars: (1) the thin disc stars which were 
distributed in the Galactic plane with a vertical scale height of about 300 pc, and (2) the thick disc stars which were distributed with a scale 
height of approximately 900 pc \citep{gilm83,juri08}. Typically, the structural thin and thick disc stars have different characteristics in 
terms of chemistry and kinematics \citep[see e.g.][]{fuhr98,felt03,soub03,bens03,holm09,lee11,adib13,bens14}. Stars also appear in two distinct 
sequences in the $[{\rm Fe}/{\rm H}]-[\alpha/{\rm Fe}]$ plane: the high-$\alpha$ sequence is traditionally called chemical thick disc while 
the low-$\alpha$ sequence is dubbed chemical thin disc \citep[see e.g.][]{reci14,nide14,hayd15,miko17}. There is a clear age difference between 
the high- and low-$\alpha$ sequence stars, the former being older than the later, as demonstrated by e.g., \citet{silv18} using the {\it Kepler} 
asteroseismic data. It should be noted that the structural, chemical and kinematic separations of the thin and thick discs are not the same thing 
\citep[see e.g.][]{hayd17}. In fact, \citet{hayd15} showed that stars belonging to the low-$\alpha$ sequence lie high above the disc midplane 
outside the solar radius, which was later explained as the flaring of mono-age populations by \citet{minc15}. The latter work explained a number 
of seemingly contradictory observations regarding thick disc formation by introducing nested flares of mono-age stellar populations. Furthermore, 
it predicted a strong age gradient in the structural thick disc, which was verified using APOGEE data by \citet{mart16}. Such complex signatures 
in the spectroscopic, kinematic and asteroseismic data can be used to unravel the formation and evolution history of the Milky Way discs through 
Galactic archaeology \citep[see][]{free02,rix13,blan16}. 

Over the past couple of decades, the computational advancement in terms of resources and numerical methods have led to the development of 
realistic hydrodynamical simulations of the formation of disc galaxies in the full cosmological context 
\citep[see e.g.][]{okam05,broo11,gued11,aume13,stin13a,mari14,wang15,gran17,naab17}. These Milky Way-mass simulations can qualitatively reproduce a 
number of observables for the Milky Way, such as flat rotation curve, disc scalelength, star formation rates (both the histories and present day 
values), structural thin and thick disc components' scale heights, as well as resolving morphological features like spiral arms and bars. The high- 
and low-$\alpha$ disc dichotomy has also been observed in cosmological simulations, and various possible formation scenarios have been discussed 
\citep[see e.g.][]{broo12,gran18a,buck20,ager20}. 

Although there have been qualitative comparisons between a few selected observables of the Milky Way with corresponding predictions of cosmological 
simulations \citep[see e.g.][]{hous11,stin13b,gome16,minc17,fatt19,frag20,gran20}, a systematic study involving observations from a large stellar 
survey is challenging primarily because of the involved selection functions associated with the observing instruments. A notable work in this 
direction was carried out by \citet{ande17} using the CoRoT \citep{bagl06b,mich08} and APOGEE data and chemodynamical model 
\citep{minc13,minc14,ande16}. They made an attempt to take into account the selection function by simply selecting stars randomly from small boxes 
in the color-magnitude diagram. In this work, we go beyond the study by \citet{ande17} by including distances along with color and magnitude in our 
selection function. Moreover, we demonstrate how closely we reproduce the observed selection function. In other words, we develop a 
framework -- taking into account proper selection function -- in which we can systematically compare any observed/inferred set of stellar properties 
coming from e.g., spectroscopic, photometric, asteroseismic, and astrometric surveys with the predictions of the Auriga suite of cosmological 
simulations by \citet{gran17}. This enables us to identify the parameter-specific discrepancies between the observation and simulations, and hence 
systematically point out the model aspects that need amendments. 

Stars are known to move radially in and out in the Galactic disc due to angular momentum transfer and scatterings. \citet{fran18,fran20} used 
parametrized models together with APOGEE data to estimate radial stellar migration. \citet{minc18} used age and metallicity from HARPS data to find
the birth places of stars. Thanks to the added dimension of precise asteroseismic ages, we demonstrate how we can use the observations in this 
framework to constrain the extent of radial migration of stars in the solar cylinder.

Since cosmological simulations are computationally expensive, we produce a lower resolution model variation for one of the halo in the Auriga 
sample to illustrate how the present study can be extended in the future to help constrain and improve galaxy formation models. This whole 
framework can be used in the future as a test-bed for any larger suite of cosmological simulations with different subgrid galaxy formation models 
for turbulence, star formation and feedback processes to unravel the formation history of the Milky Way.

\section{The observed sample}
\label{sample}
\begin{figure}
\includegraphics[scale=0.4]{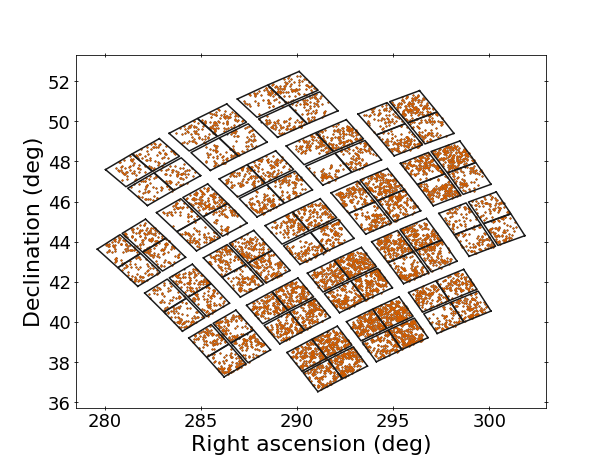}
\caption{Spatial distributions of stars in the {\it Kepler} field.}
\label{fig1}
\end{figure}

We have a total of 7186 targets in the observed sample used in this study. These stars were all observed by the {\it Kepler} spacecraft during its 
nominal mission \citep{gill10b,koch10}. In Figure~\ref{fig1}, the targets are shown in the fixed {\it Kepler} field-of-view. We retrieved the global 
asteroseismic quantities, namely the frequency of maximum power, $\nu_{\rm max}$, and the large frequency separation, $\Delta\nu$, from \citet{yu18}. 
We used the solutions found by the \textsc{SYD} pipeline \citep{hube09} to be consistent with the solar values of $\nu_{\rm max}$ and $\Delta\nu$ 
used in the stellar properties determination (see further details below).

All stars have chemical abundances available from the 16th data release of the Apache Point Observatory Galactic Evolution Experiment 
\citep[APOGEE DR16;][]{apog17,ahum20}, along with measured $JHK_s$ band photometric magnitudes from the Two Micron All-Sky Survey 
\citep[2MASS;][]{skru06}. We also have six-dimensional astrometric measurements (celestial positions, proper motions, parallax and line-of-sight 
velocity) for all stars in our sample from the second data release of {\it Gaia} \citep[{\it Gaia} DR2;][]{gaia18}. The systematic zero 
point offset for the {\it Gaia} DR2 parallaxes \citep{lind18} was corrected following \citet{zinn19a} as their correction was computed based on 
similar stars within the {\it Kepler} field-of-view.

Besides requiring the availability of the above information for all stars, we also had the following quality criteria applied in this work:
\begin{itemize}
\item {The 2MASS quality flag \texttt{Qflg} $=A$ for all bands.}
\item {No \texttt{TEFF\_BAD} or \texttt{VMICRO\_BAD} ASPCAP flag set in APOGEE.}
\item {No \texttt{LOW\_SNR}, \texttt{PERSIST\_JUMP\_POS}, \texttt{PERSIST\_JUMP\_NEG}, \texttt{SUSPECT\_RV\_COMBINATION}, 
\texttt{VERY\_BRIGHT\_NEIGHBOR} or \texttt{PERSIST\_HIGH} star flag set in APOGEE.}
\item {$\varpi > 0$ and $\varpi / \sigma_{\varpi} > 5$, where $\varpi$ and $\sigma_{\varpi}$ denote the parallax and its uncertainty, respectively.}
\end{itemize}

We note that removing the positive parallaxes with inverse fractional uncertainties above a threshold as well as negative parallaxes can bias the 
sample \citep{luri18}. However, since we compare it with simulation samples cut in the same way, this bias is not an issue in this study.

The stellar properties, including ages for the sample were determined using the BAyesian STellar Algorithm \citep[\textsc{BASTA};][]{silv15,silv17}. 
Briefly, \textsc{BASTA} uses a Bayesian approach and computes the probability density functions for stellar properties by combining a set of stellar 
models with the observational constraints and prior information. In this study, we used the BaSTI (a Bag of Stellar Tracks and Isochrones) stellar 
models and isochrones library \citep{Hida18} with overshooting and no mass-loss. For the reference solar values, we adopted 
$\nu_{{\rm max,}\odot} = 3090$ $\mu$Hz, $\Delta\nu_{\odot} = 135.1$ $\mu$Hz \citep{hube11}, and effective temperature $T_{{\rm eff,}\odot} = 5777$ K. 
We used the Salpeter Initial Mass 
Function \citep{salp55} as a prior to quantify our expectation of mostly low-mass stars. Furthermore, we constrained the ages of our stars to be 
within the age of the Universe (13.6 Gyr; to be consistent with the simulations). For all stars in our sample, we fitted the effective temperature 
$T_{\rm eff}$, metallicity $[{\rm Fe}/{\rm H}]$, large frequency separation $\Delta\nu$ (corrected following the prescription of \citet{sere17}), 
frequency of maximum power $\nu_{\rm max}$, the corrected parallax $\varpi$ along with apparent magnitudes in the 2MASS $JHK_s$ photometric bands 
using the \citet{gree19} dust map to account for extinction. In addition, asteroseismology can inform us whether a red giant star is burning hydrogen 
in the shell or helium in the core \citep[see e.g.][]{bedd11}. We used the evolutionary phase information available from asteroseismology as a 
Bayesian prior in determining the stellar properties. The resulting stellar ages have a median uncertainty of about 16\% (which is a conservative 
estimate because we symmetrized the errorbar by using the maximum of the negative and positive uncertainties).

\section{The Auriga simulations and mock catalogues}
\label{catalogue}
The Auriga project\footnote{https://wwwmpa.mpa-garching.mpg.de/auriga/} includes a suite of 30 high-resolution cosmological magneto-hydrodynamical 
zoom-in simulations of the formation of Milky Way-mass galaxies \citep[see][]{gran17}. Following the naming convention used in the original paper, 
we shall refer the simulations as Au01, Au02, $\ldots$, Au30 in the present study. The simulations were performed using a moving-mesh, N-body, 
magneto-hydrodynamics code \citep[AREPO;][]{spri10}. We refer the reader to the paper by \citet{gran17} for further details on the model physics 
considered as well as on the physical properties of all the 30 simulated galaxies. Briefly, the computation of an Auriga zoom simulation involves 
the following steps.
\begin{enumerate}
\item Identify an isolated host dark matter halo of virial mass lying within the range $[1 \times 10^{12}, 2 \times 10^{12}]$ M$_\odot$ at 
the redshift $z = 0$ from the Eagle dark matter only simulation by \citet{scha15}.
\item Trace the halo back in time at $z = 127$. Subsequently, split the dark matter into the dark and baryonic components according to the measured
parameters of the standard $\Lambda$CDM model of Big Bang cosmology \citep{plan14}.
\item Increase the resolution of the Lagrangian region around the halo, and degrade the resolution of distant particles.
\item Evolve it back to $z = 0$ using the AREPO code with a comprehensive galaxy formation model. 
\end{enumerate}

Despite a significant advancement in computational resources during the past a couple of decades, current hydrodynamical simulations of galaxy
formation can not resolve individual stars. The mass resolutions of the dark and baryonic matter particles in the Auriga simulations are about 
$4 \times 10^4$ and $5 \times 10^3$ M$_\odot$, respectively. A star particle in the simulation represents a single stellar population of a given 
age and metallicity. Following two different approaches (as described below), 
\citet{gran18b} presented mock {\it Gaia} DR2 stellar catalogues for six simulations (Au06, Au16, Au21, Au23, Au24 and Au27) that had distinct 
properties. Specifically, they selected Au16 and Au24 for their large discs; Au06 for being a close Milky Way analogue in terms of stellar mass, 
star formation rate, thin and thick discs' scale heights, and morphology; and Au21, Au23 and Au27 for their interesting satellite interactions.

The process of generating mock catalogues requires specification of the solar position and velocity. \citet{gran18b} presented four mock catalogues 
for each of the above six simulations with four different solar positions spread at equidistant azimuthal angles. They chose their reference solar 
azimuth at 30 deg behind the major axis of the bar (the corresponding catalogue is referred to as the reference catalogue), while other three 
positions were assumed at 120, 210 and 300 deg behind the bar. The distance between the Sun and Galactic centre was assumed to be 8 kpc \citep{reid93}, 
and the Sun's height above the Galactic plane, 20 pc \citep{hump95}. They assumed the solar velocity with respect to the local standard of rest to 
be [11.1, 12.24, 7.25] km s$^{-1}$ \citep{scho10}. \citet{gran18b} created mock catalogues following two different approaches: (1) using the 
parallelized version of SNAPDRAGONS code \citep[HITS-MOCKS;][]{hunt15}, and (2) using the method described in \citet[ICC-MOCKS;][]{lowi15}. This 
means that we have a total of eight catalogues for each simulation. In this study, we shall use the ICC-MOCKS as reference, whereas the HITS-MOCKS 
for Au06 will be used to assess the robustness of our conclusions against the uncertainties associated with methods used to generate mock 
catalogues. We refer the reader to \citet{gran18b} for a detailed discussion on the limitations of the two mocks as well as advantages of one over 
the other. 

The different mock catalogues for all the six simulations are publicly available\footnote{http://dataweb.cosma.dur.ac.uk:8080/gaia-mocks/}. 
The catalogues provide several quantities including astrometric and photometric parameters, radial velocity, stellar properties and their 
uncertainties. A complete list of parameters contained in the catalogues can be found in Table~A1 of \citet{gran18b}. The mock catalogues also 
contain unique ID of the parent simulation particle, which can be used to extract more information about star particles -- such as elemental 
abundances, birth position and birth velocity -- from the snapshots.

We analyse also a model variation for Au06 which includes: i) a softer equation of state (EOS) for the subgrid interstellar medium (ISM) model 
\citep{voge13}; ii) a metallicity-dependent wind velocity prescription \citep[see e.g., equation 3 of][]{pill18}; and iii) a thermal and kinetic 
active galactic nuclei (AGN) jet feedback model \citep{wein17}. The soft EOS interpolates between an isothermal equation of state and the full 
\citet{spri03} effective equation of state, with an interpolation parameter $q=0.3$. This describes a case in which a fraction of the stellar feedback 
energy is retained by the ISM, resulting in a less pressurised ISM relative to the fiducial case. The metal-dependent wind scheme adopts a scaling 
that relates the velocity of galactic winds with gas metallicity, motivated by additional radiative losses implied by higher metallicity galaxies 
\citep[e.g.][]{scha15}. This simulation is lower in resolution by a factor of about 8 in mass, and the corresponding HITS mock catalogue was generated 
using the SNAPDRAGONS code \citep{hunt15}. Hereafter, we refer to this simulation as soft EOS to distinguish it from the fiducial Auriga model 
simulations.

\begin{figure*}
\includegraphics[scale=1]{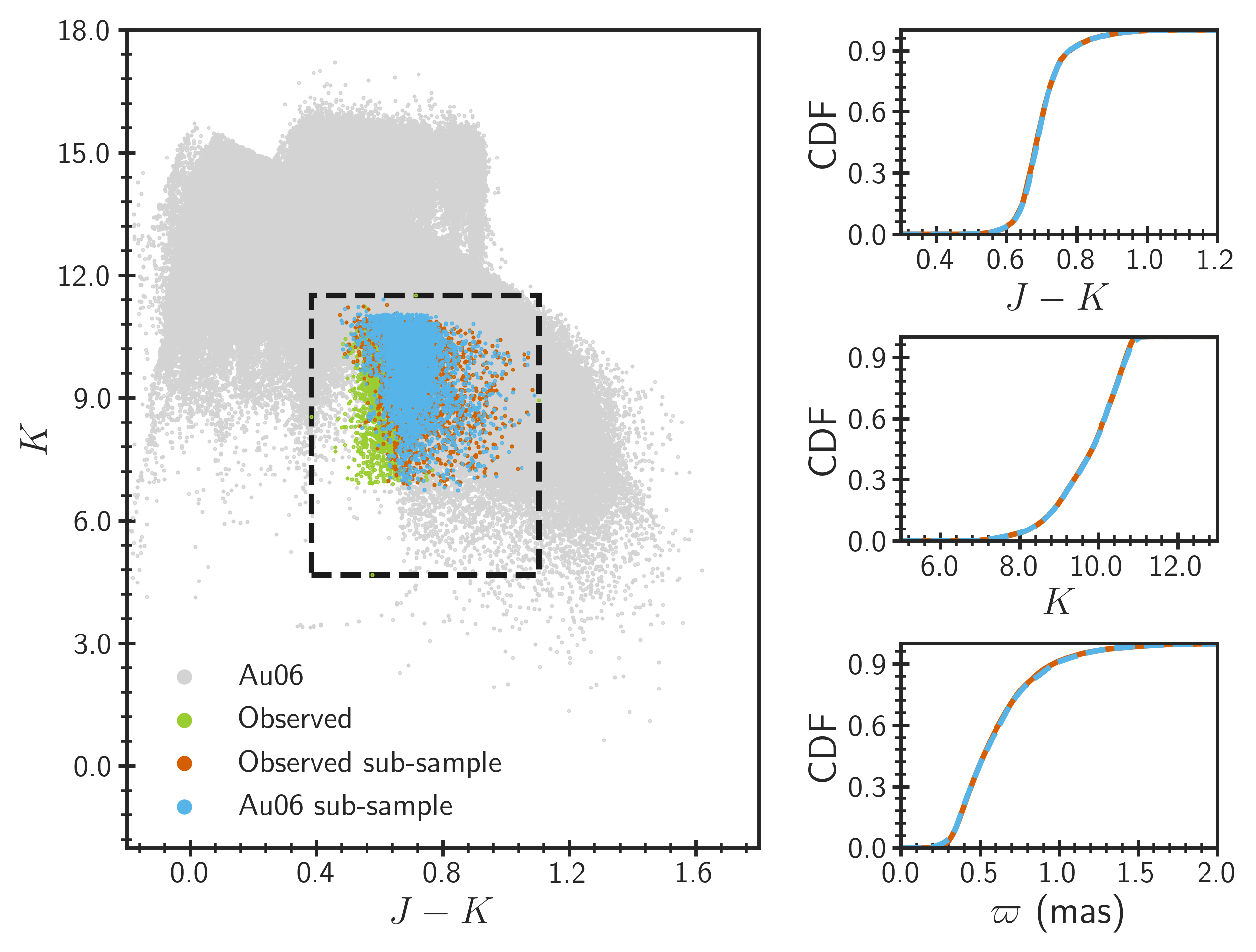}
\caption{Illustration of the selection function using the Au06 ICC catalogue generated assuming the standard solar azimuth. The left panel shows 
the colour-magnitude diagram for Au06 stars in the {\it Kepler} field-of-view (grey dots), observed stars (green dots), sub-sampled observed stars 
(orange dots) and sub-sampled Au06 stars (blue dots). The rectangular box encloses all the observed stars. The right panels demonstrate the 
cumulative distribution functions of the sub-sampled observed (orange solid curve) and sub-sampled Au06 stars (blue dashed curve) for the colour 
(top panel), magnitude (middle panel) and the parallax (bottom panel). See the corresponding text for details.}
\label{fig2}
\end{figure*}

\section{The selection function}
\label{selection}
In this study, we aim to perform direct comparisons between observations and simulations. However, before we can do such comparisons sensibly 
we need to carefully account for the biases in the observations. The observed sample used in the current study is biased mainly for three reasons. 
First, unlike simulations the dataset is spatially limited to the {\it Kepler} field-of-view (see Figure~\ref{fig1}). 
This is straight-forward to deal with because we can easily extract the stars from simulations that are in the {\it Kepler} field-of-view 
and perform the comparisons. Secondly, the sample is limited due to the Malmquist bias (or selection bias), i.e. only intrinsically brighter stars 
can be observed at larger distances. Finally, the observed sample is biased because the {\it Kepler} stars were selected based on a number of 
heterogeneous criteria \citep{hube10,pins14}. The combined effect of the last two is substantially more complicated.

A careful account of such biases is also necessary when studying the structural properties of the Milky Way using observed stellar 
populations. For example, \citet{casa16} used photometry to derive the colour and magnitude limits within which their asteroseismic 
sample was representative of an underlying unbiased photometric sample. Moreover, they used several different approaches, including population 
synthesis and Galaxy modelling, to account for the {\it Kepler} target selection. We note that the inferences of physical properties of the Milky 
Way, as done in the above studies, ideally require observation of the true underlying stellar populations. However, this is not the case for our 
specific problem at hand: we can make meaningful comparisons between the observation and simulations and draw conclusions even if they are biased 
(as long as both are biased in the same way).

In this study, we take a slightly different approach, in which we sub-sample simulation stars based on the observation, ensuring similar biases 
in both. This is done by first extracting all the simulation stars from the {\it Kepler} field-of-view. Similar to the observation, 
we consider only those simulation stars that have reliable parallaxes, i.e. the stars with positive parallaxes and with their inverse 
fractional uncertainties greater than 5. In the colour-magnitude diagram shown in the left panel of Figure~\ref{fig2}, these stars are shown in 
grey for the reference ICC catalogue for Au06. To ensure similar biases, we sub-sample stars from both the observation and simulation in such a 
way that their distributions of $J - K$ colour, $K$ magnitude and $\varpi$ are similar. This process is explained below in more detail using 
example of the reference ICC catalogue for Au06. 

To improve the computational efficiency, we first discard all the Au06 stars outside the observed ranges in $J - K$ colour, $K$ magnitude and 
parallax. For simplicity, this is illustrated in the projected 2D space in the left panel of Figure~\ref{fig2}. We get rid of the Au06 stars 
outside the rectangular box, which encloses all the observed stars as shown with the green dots. Subsequently, we discretize the resulting 
datacube with $n_{J - K} + 1$, $n_K + 1$ and $n_{\varpi} + 1$ uniformly spaced points along the $J - K$, $K$ and $\varpi$ dimensions, respectively. 
This divides the full observed space into $n_{J - K} \times n_K \times n_{\varpi}$ small sub-spaces (or cells). For each cell, we randomly choose 
as many Au06 stars as there are observed in that cell. Note that a simulation of the Milky Way should ideally produce a number of stars in each 
cell which is greater than or equal to the number of observed stars in that cell. In reality however, the models of galaxy formation are uncertain, 
resulting in fewer simulation stars in some cells compared to the observation. For such cells, we randomly choose as many observed stars as there 
are Au06 stars in that cell, i.e. disregard some of the observed stars as well.

By construction, the above process results in two sub-samples -- one of the observed sample and the other of Au06 mock catalogue -- of same size. 
Clearly, the sub-samples also have similar distributions in $J - K$ colour, $K$ magnitude and parallax. The similarities of the distributions 
depend on the choice of the number of points along each dimension: the larger the number of points, the more similar the distributions. Since the 
choice of arbitrarily large numbers for $n_{J - K}$, $n_K$ and $n_{\varpi}$ reduces sub-samples' size significantly, we find their optimal values 
through an iterative process, which involves the following steps.
\begin{enumerate}
\item Initialize $n_{J - K} = 5$, $n_K = 5$ and $n_{\varpi} = 5$.
\item Find the sub-samples of the observed sample and Au06 mock catalogue following the procedure described above.
\item Assess similarities of the distributions of $J - K$ colour, $K$ magnitude and $\varpi$ from the two sub-samples using K-sample 
Anderson-Darling test. 
\item Increase the number of points along the dimensions by 5 if the corresponding null hypotheses -- that the two sub-samples are drawn from the 
same distribution -- are rejected at the 5\% level, and repeat from step (ii). 
\end{enumerate} 
Following the above process, we found that the null hypotheses could not be rejected for $n_{J - K} = 30$, $n_K = 25$ and $n_{\varpi} = 40$. The 
corresponding observed and Au06 sub-samples are shown in orange and blue dots in the left panel of Figure~\ref{fig2}, and the cumulative 
distribution functions (CDFs) for the colour, magnitude and parallax are compared in the right panels.

\begin{figure}
\includegraphics[scale=0.5]{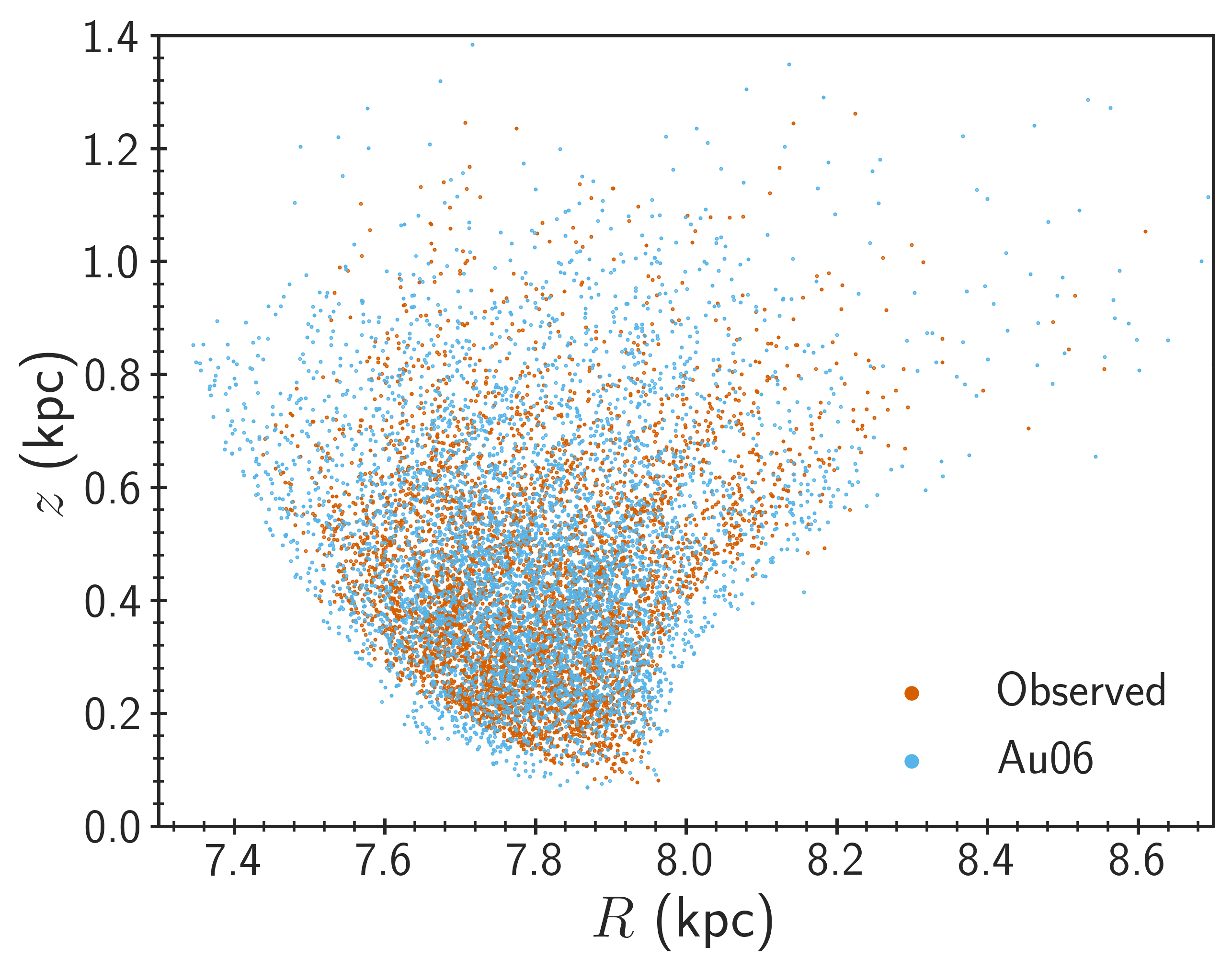}
\caption{Spatial distributions of the observed (orange) and Au06 (blue) stars after the selection function.}
\label{fig3}
\end{figure}

\begin{figure}
\includegraphics[scale=0.5]{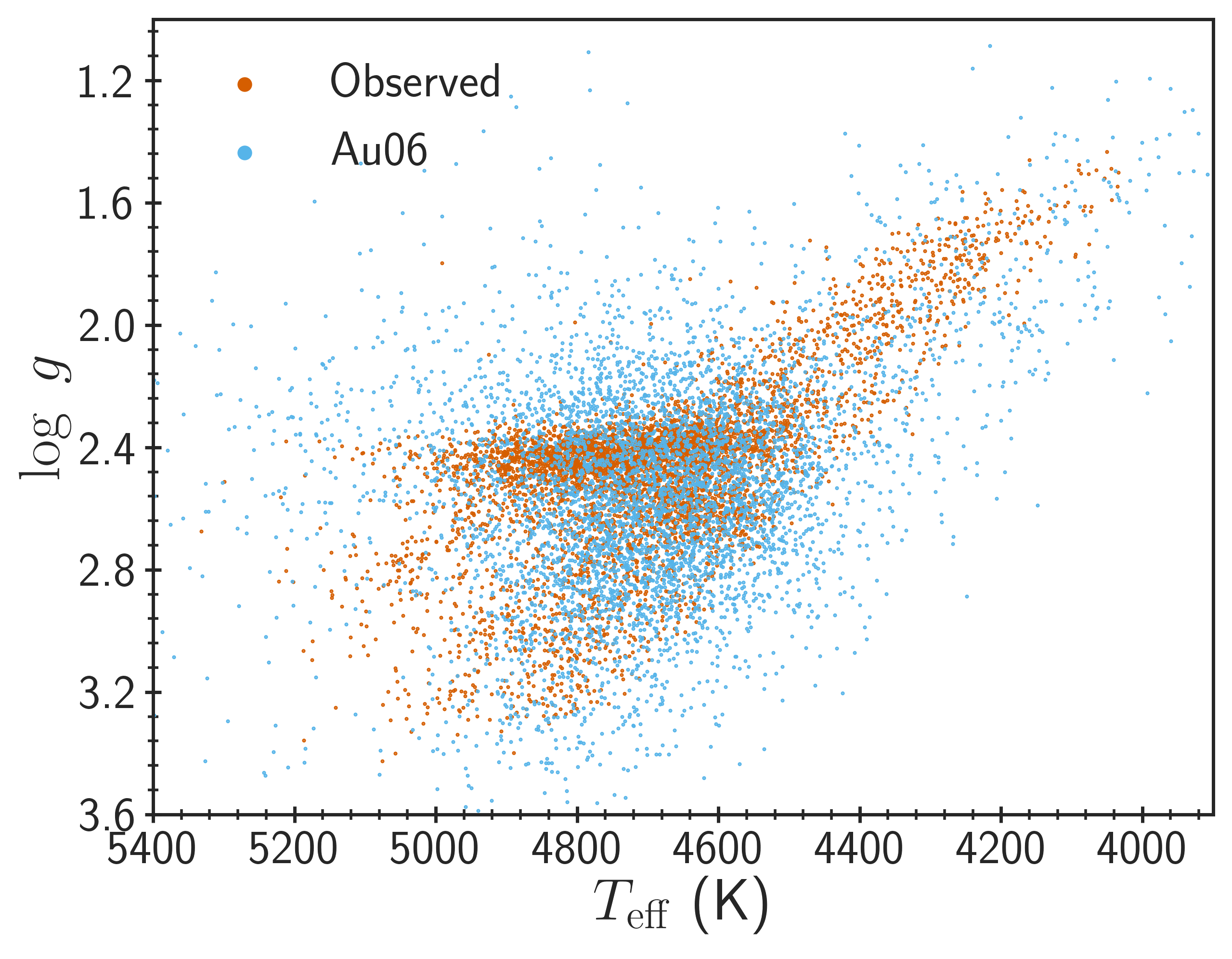}
\caption{Kiel diagrams for the observed (orange) and Au06 (blue) stars after the selection function.}
\label{fig4}
\end{figure}

\begin{table*}
\centering
\scriptsize
\caption{The final sample size after selection function and the estimated radial stellar migration. The columns are divided into four groups based on the solar azimuths assumed in generating mock catalogues. The size is the number of stars in the sample. The $\Delta{R}_{\textless{4}}$ and $\Delta{R}_{\in [4,8]}$ denote radial migrations for young ($< 4$ Gyr) and old ($\in [4, 8]$ Gyr) stellar populations, respectively.}
\label{tab1}
\begin{tabular}{cccccccccccccccc}
\hline\hline
            &             \multicolumn{3}{c}{The Sun at 30 deg}                        & &               \multicolumn{3}{c}{The Sun at 120 deg}                     & &            \multicolumn{3}{c}{The Sun at 210 deg}                        & &            \multicolumn{3}{c}{The Sun at 300 deg}                       \\
\cline{2-4} \cline{6-8} \cline{10-12} \cline{14-16} \\
Simulation  &  size  &  $\Delta{R}_{\textless{4}}$     &  $\Delta{R}_{\in [4,8]}$  & &  size  &  $\Delta{R}_{\textless{4}}$     &  $\Delta{R}_{\in [4,8]}$  & &  size  &  $\Delta{R}_{\textless{4}}$     &  $\Delta{R}_{\in [4,8]}$  & &  size  &  $\Delta{R}_{\textless{4}}$     &  $\Delta{R}_{\in [4,8]}$ \\
            &        &             (kpc)               &         (kpc)             & &        &            (kpc)                &          (kpc)            & &        &              (kpc)              &           (kpc)           & &        &               (kpc)             &           (kpc)          \\
\hline
Au06(ICC)   &  5216  &               1.99              &          2.03             & &  4782  &              1.42               &            2.09           & &  5143  &               1.94              &             2.06          & &  5453  &                 1.17            &            1.91          \\
Au16(ICC)   &  4299  &               1.05              &          3.09             & &  4478  &              1.32               &            3.11           & &  4262  &               1.11              &             3.35          & &  4270  &                 0.86            &            3.13          \\
Au21(ICC)   &  5510  &               1.82              &          2.92             & &  5456  &              1.90               &            3.04           & &  5396  &               1.95              &             2.81          & &  5679  &                 2.21            &            2.87          \\
Au23(ICC)   &  5246  &               1.69              &          2.74             & &  5868  &              1.78               &            2.45           & &  5334  &               2.12              &             2.32          & &  5313  &                 1.68            &            2.57          \\
Au24(ICC)   &  4455  &               2.08              &          3.52             & &  4675  &              1.38               &            3.32           & &  4663  &               1.13              &             3.70          & &  4354  &                 1.54            &            3.53          \\
Au27(ICC)   &  6008  &               1.66              &          2.43             & &  5769  &              1.31               &            2.36           & &  5524  &               1.24              &             2.25          & &  5636  &                 1.08            &            2.24          \\
Au06(HITS)  &  5805  &               1.99              &          2.03             & &  4995  &              1.23               &            2.17           & &  5777  &               1.75              &             2.13          & &  6165  &                 1.22            &            1.99          \\
\hline
\end{tabular}
\end{table*}

In Figure~\ref{fig3}, we show the spatial distributions of stars in the observed and Au06 sub-samples. As expected, both the observed and 
simulation stars are from the same region in the Milky Way. Moreover, there is a reasonable agreement between the distributions of the observed 
and simulation stars in the $R-z$ plane. Figure~\ref{fig4} compares the two sub-samples in the Kiel diagram. Clearly, the stars in the two 
sub-samples have similar spectral type. The above ensures that we can now use the two sub-samples, and make meaningful comparisons. It should be 
noted that we go through the above process before comparing any of the simulation mock catalogues with the observation, which means we shall 
implicitly work with sub-samples in the subsequent sections (unless stated explicitly otherwise). On average, our resulting sub-samples 
contain about 5200 stars compared to the original 7186 observed targets.

\section{Results}
\label{results}
We applied the selection function described in Section~\ref{selection} to all four ICC catalogues generated assuming solar azimuths at 
30, 120, 210 and 300 deg behind the major axis of the bar for each of the six simulations (Au06, Au16, Au21, Au23, Au24 and Au27), as well as to 
the all four HITS catalogues generated for the Au06 simulation. This results in a total of 28 pairs of samples (one pair corresponding to each mock 
catalogue). The final sample size after the selection function for all pairs is listed in Table~\ref{tab1}, which varies from 4262 to 6165 depending 
on the simulation and mock catalogue. In the next few sub-sections, we shall compare the predictions of simulations with the observation.

\begin{figure*}
\includegraphics[scale=1]{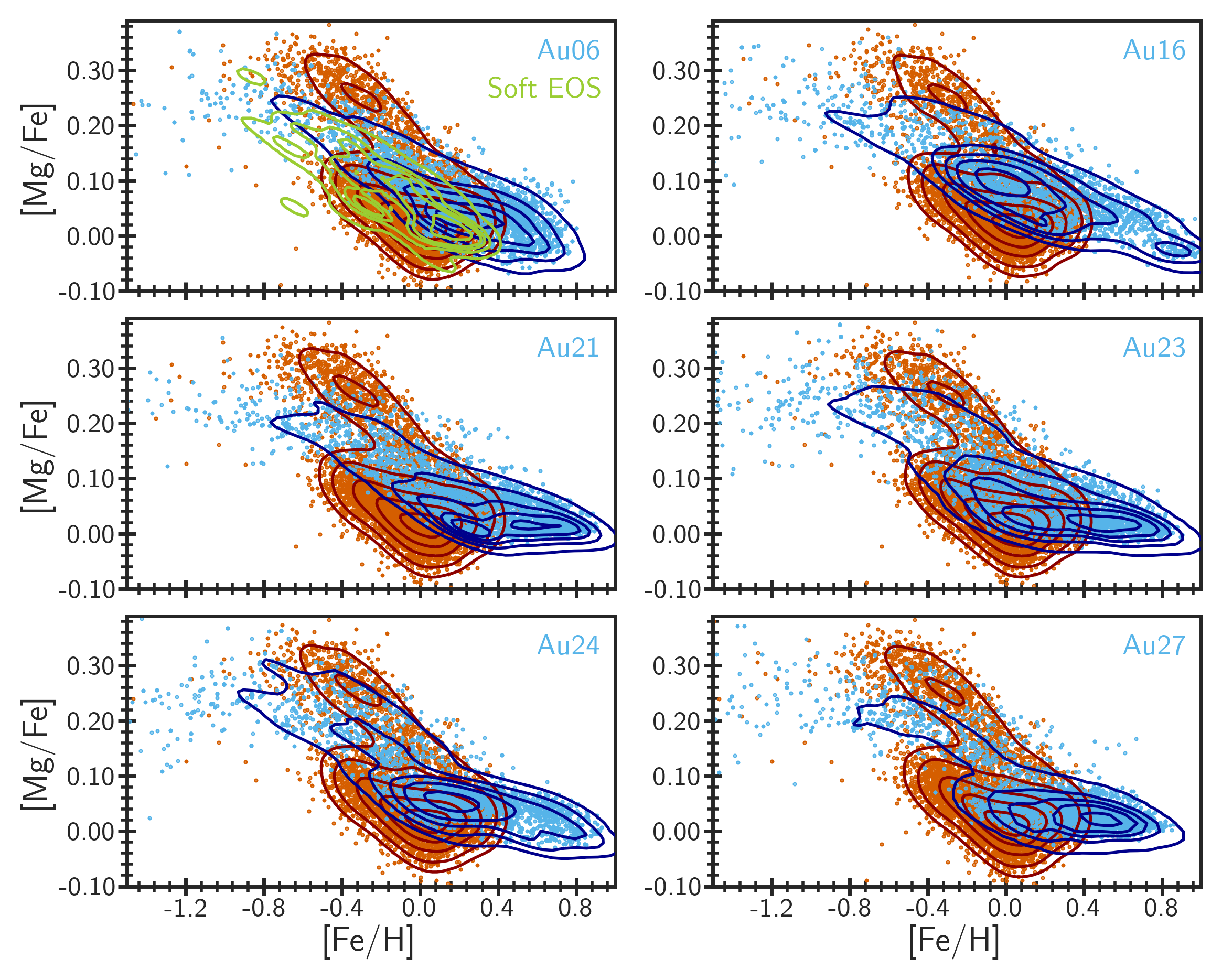}
\caption{Comparison between the observed and simulation abundance distributions in $[{\rm Fe}/{\rm H}]-[{\rm Mg}/{\rm Fe}]$ plane. The six different 
panels correspond to different simulations with the respective ICC catalogue generated assuming the standard solar azimuth. In each panel, the 
orange and blue dots represent the observed and simulation stars, respectively, while the five dark red (blue) innermost-to-outermost contours 
enclose approximately 10, 30, 50, 70 and 90\% of the observed (simulation) stars. In top left panel, the green contours show the model variation
for Au06 as discussed in Section~\ref{catalogue}. The corresponding dots are not shown for the sake of clarity.}
\label{fig5}
\end{figure*}

\subsection{Comparison of the abundance distributions in $[{\rm Fe}/{\rm H}]-[\alpha/{\rm Fe}]$ plane}
\label{abundances}
The $\alpha$-elements primarily form from core collapse supernovae on Myr time scales while iron forms mainly from Type Ia supernovae on Gyr time
scales \citep[see e.g.][]{matt86}. This makes the abundance plane $[{\rm Fe}/{\rm H}]-[\alpha/{\rm Fe}]$ particularly interesting: it is now well 
known that this plane contains key information about the formation and evolution of the Galactic discs \citep[see e.g.][]{matt12,andr17}. For instance, 
the observed high- and low-$\alpha$ sequences in this plane can be explained by the so-called two-infall chemical evolution 
models \citep[see e.g.][]{spit19,spit20,spit21}, which hypothesized that there were two episodes of gas accretion at two different epochs (well 
separated in time) in the formation history of the Milky Way. On the other hand, a similar distribution in the abundance plane can be produced with  
analytical chemodynamical models including effects of radial migration and kinematic heating, as recently shown by e.g., \citet{shar20}.

In Figure~\ref{fig5}, we compare the abundance predictions from the Auriga simulations with the observation in $[{\rm Fe}/{\rm H}]-[{\rm Mg}/{\rm Fe}]$ 
plane. We assume the $[{\rm Mg}/{\rm Fe}]$ abundance ratio to be the tracer of $\alpha$-element abundances. It is well known that simulations using 
yields from \citet{port98}, as is the case for the Auriga simulations, underproduce magnesium by a factor of about 2.5 
\citep[see e.g.][]{voor20}. Therefore, we corrected the magnesium abundance by this factor. We wish to point out that, since magnesium is not a strong 
coolant, its underproduction is not expected to affect the dynamics of simulations. We note that distributions of the observed stars 
differ slightly from one panel to another because of the selection function (see Section~\ref{selection}). A dichotomy can easily be seen in the 
observed abundances. We can see a variety of predicted distributions in different panels, some of which also show a dichotomy with different 
morphology. The predicted dichotomy for Au27 is particularly similar to the observed one. Qualitatively, the results are similar for the HITS 
catalogues (see Figure~\ref{afig1}) and for the catalogues generated assuming different solar azimuths (see Figures~\ref{afig2}, \ref{afig3} and 
\ref{afig4}), although some differences may also be noticed. 

Although some simulations do produce abundance dichotomy, there are several differences between the distributions of the observed and 
simulated abundances. For instance, all simulations seem to predict systematically larger $[{\rm Fe}/{\rm H}]$ compared to the observation. 
Moreover, there are clear morphological differences between the observed dichotomy and the closest matching dichotomy for Au27. These differences 
are likely related to the used input physics in the Auriga simulations, in particular they are expected to depend on the assumed yields and 
feedback processes. The suite of Auriga simulations used yields from \citet{kara10} for asymptotic giant branch stars and from \citet{port98} for 
core collapse supernovae. Certain features of the abundance distribution predicted by the recent VINTERGATAN simulation, which uses the yields from 
\citet{woos07}, appear to agree slightly better with the observations. For instance, their predicted $[{\rm Fe}/{\rm H}]$ is less than 0.6 
\citep[see Figure~15 of][]{ager20}. However, they have similar issue with the oxygen abundance, which is overproduced in the simulation.

\begin{figure*}
\includegraphics[scale=1]{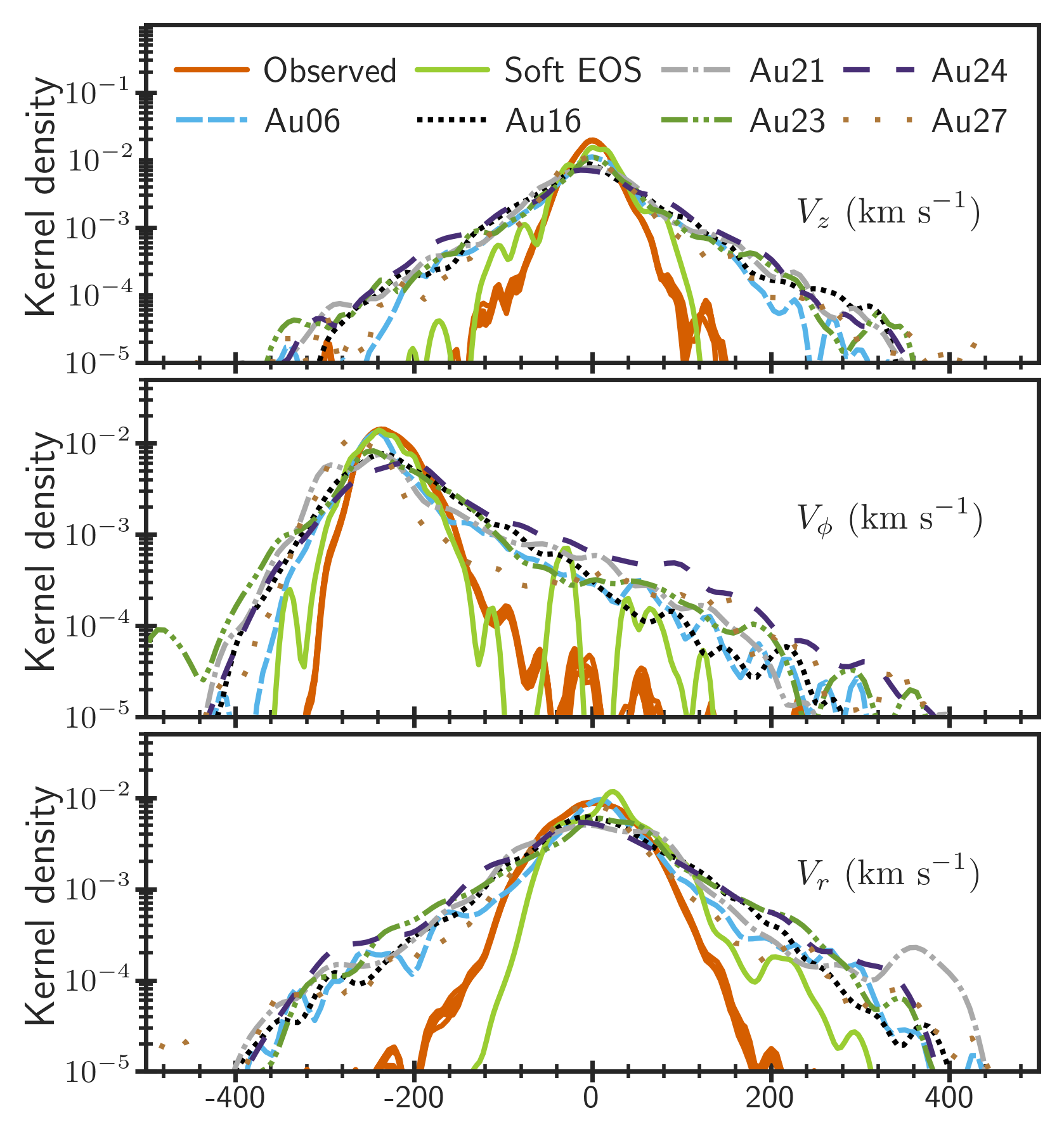}
\caption{Comparison between the observed and simulation Galactocentric velocity distributions. The top, middle and bottom panels show the vertical, 
azimuthal and the radial components of the velocity distributions in cylindrical coordinate, respectively. In each panel, solid orange curves 
represent observed samples (see the corresponding text for details), while the other curves correspond to the six simulations (see the legend in the top 
panel) with the respective ICC catalogue generated assuming the standard solar azimuth. The solid green curves show the model variation
for Au06 as discussed in Section~\ref{catalogue}.}
\label{fig6}
\end{figure*}

To test if we can improve our simulations, we produced a model variation for Au06 as discussed in Section~\ref{catalogue}. As shown in the top 
left panel of Figure~\ref{fig5}, this model avoids the issue of large metallicity predicted by the fiducial Auriga simulations. We can use our framework 
in the future for new simulations with yields from different sources and with various subgrid models for feedback processes \citep[e.g.][]{buck20,ager20} 
to better understand the chemical evolution of the Milky Way in full cosmological context.

\subsection{Comparison of the velocity distributions}
\label{velocities}
We used the 6D phase-space information of the observed sample and simulation mock catalogues to compute their Galactocentric velocity in 
cylindrical coordinates using {\tt Astropy}\footnote{https://www.astropy.org} \citep{astr13,astr18}. To be consistent, we used the same solar
position and velocity in the transformation as used by \citet{gran18b} in generating the mock catalogues.

In Figure~\ref{fig6}, we compare velocity distributions predicted by the Auriga simulations with the corresponding distributions of the observed 
samples. Recall from Section~\ref{selection} that we get a slightly different observed sample due to the selection function depending on the simulation 
mock catalogue at hand for comparison. In Figure~\ref{fig6}, the different orange curves (which are very similar and hence overlap substantially) show 
velocity distributions for observed sub-samples corresponding to the six simulation mock catalogues. As we can see in the figure, for all of the 
three velocity components, the peaks of the distributions for the observation and simulations agree reasonably well. However, all six simulations 
predict much longer tails in the velocity distributions compared to the observed sample. This systematic difference is likely due to inaccuracies 
in the currently used subgrid models for turbulence, star formation and feedback processes in the Auriga project, which can be tested in the 
future by comparing simulations generated using revised recipes of these physical processes (as discussed further in this section below) with the
observation. 

\begin{figure}
\includegraphics[scale=0.5]{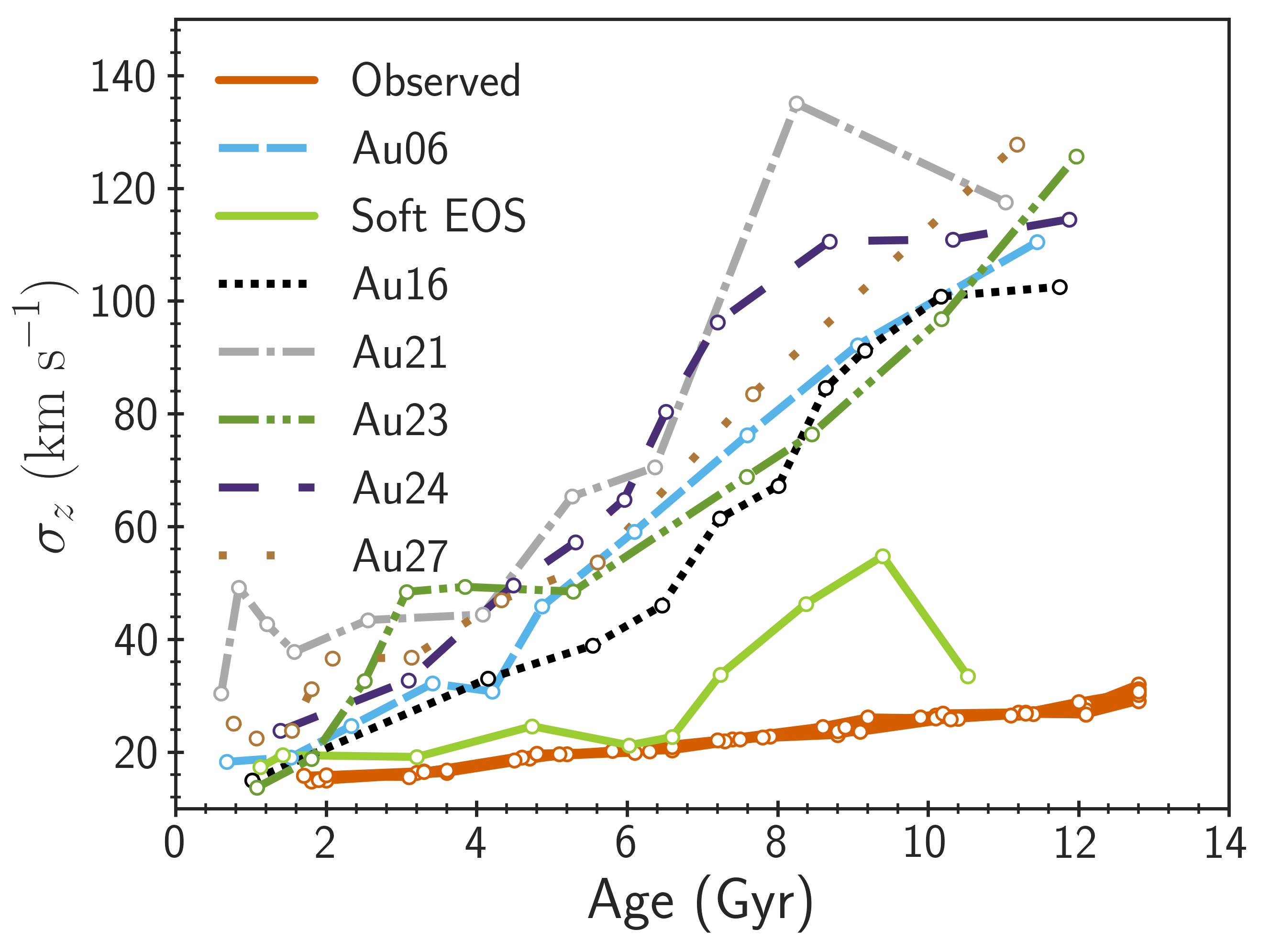}
\caption{Comparison between the observed and simulation age-velocity dispersion relation. Solid orange curves represent observed samples (see the 
corresponding text for details), while the other curves correspond to the six simulations (see the legend) with the respective ICC catalogue 
generated assuming the standard solar azimuth. The solid green curve shows the model variation for Au06 as discussed in 
Section~\ref{catalogue}. The uncertainties in the observed data were propagated to the computed velocity dispersion using Monte Carlo simulation 
(uncertainties lie within the thickness of orange curves). The age bins were chosen in such a way that they contain approximately same numbers 
of stars.}
\label{fig7}
\end{figure}

Figure~\ref{fig7} shows the vertical velocity dispersion as a function of age. We used the so-called biweight midvariance -- a robust statistic 
as implemented in {\tt Astropy} with standard tuning constant, $c = 9$ -- to determine the velocity variance (square root of which gives the velocity 
dispersion). Again, the orange curves in the figure show the velocity dispersion for all of the six observed sub-samples. As we can see in the figure, 
all of the six simulations predict substantially `hotter' kinematics compared to the Milky Way. The above results are qualitatively independent of 
the choice of mock catalogues (see Figure~\ref{afig5} for results obtained using HITS catalogues) and also the solar azimuths assumed in generating 
the mock catalogues (see Figures~\ref{afig6}, \ref{afig7} and \ref{afig8} for results obtained assuming solar azimuths 120, 210 and 300 deg, 
respectively).

This is a well known issue related to cosmological simulations. \citet{hous11} used seven cosmological simulations run with different N-body 
hydrodynamical galaxy formation codes to study the age-velocity dispersion relation (AVDR) for disc stars. They concluded that all of the analysed 
simulations predict too large velocity dispersions compared to the Milky Way disc. \citet{hous11} pointed out that the velocity dispersion of a 
simulation has a lower limit which depends on the treatment of the heating and cooling of the interstellar medium, and on the assumed density 
threshold for star formation. 

Traditionally, cosmological zoom-in simulations used small values of the star formation density threshold compared to the typically observed 
densities of star forming regions (density of a giant molecular cloud $\sim100$ cm$^{-3}$). For example, \citet{gove07,hous11} used a density 
threshold of 0.1 cm$^{-3}$. This was mainly because those simulations could not resolve the observed densities of star forming regions. The Auriga 
simulations used a density threshold of 0.13 cm$^{-3}$ \citep{gran17}, which was derived from the parameters describing the ISM 
and the desired star formation time-scale \citep{spri03}. Recently, \citet{bird20} used the high-resolution cosmological zoom-in simulation {\tt h277} 
from \citet{chri12} to closely reproduce the measured solar-neighbourhood AVDR of \citet{casa11}. They attributed this success mainly 
to the simulation's ability to form stars in dense and cold environment ($n > 100$ cm$^{-3}$ and $T < 1000$ K, where $n$ and $T$ are number 
density and temperature, respectively), similar to those in giant molecular clouds. Although these results are indeed promising, several aspects (and 
not just the density threshold) of the subgrid galaxy formation physics affect the resulting velocity dispersion. In fact, simulations have been 
carried out with a density threshold as large as 1000 cm$^{-3}$ which still result in too hot kinematics \citep{sand20}. 

Our preliminary test with softer equation of state produces velocity distributions and AVDR which are in better agreement with the observations
(compare solid green and orange curves in Figures~\ref{fig6} and \ref{fig7}). We can use our framework together with precise asteroseismic ages in the
future to distinguish between the various possible physical scenarios relevant in the formation of the Milky Way.

\begin{figure*}
\includegraphics[scale=1]{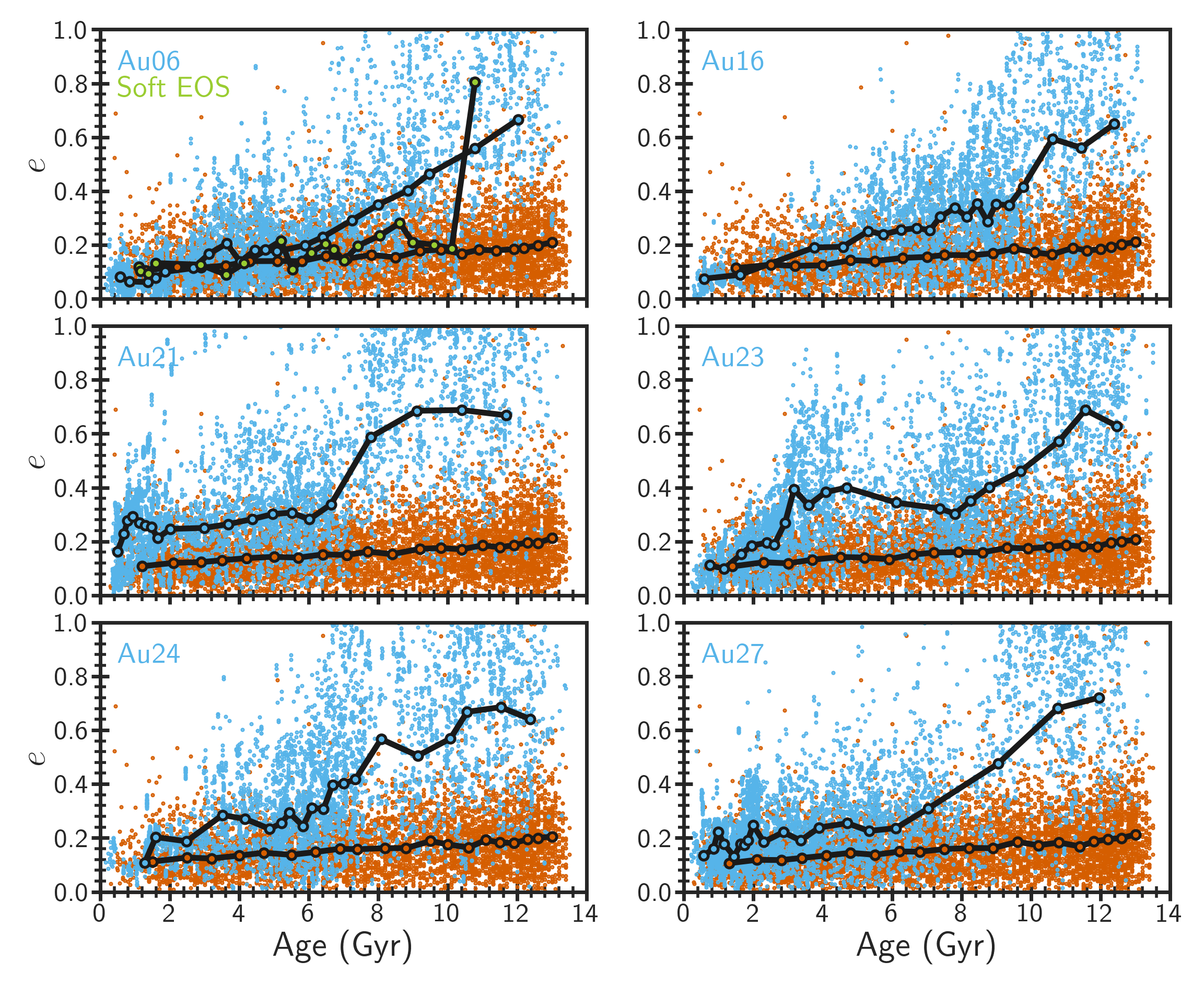}
\caption{Orbital eccentricity of the observed and simulation stars as a function of age. The six different panels correspond to different simulations (see 
the legends) with the respective ICC catalogue generated assuming the standard solar azimuth. In each panel, the orange and blue dots represent 
the observed and simulation stars, respectively. To guide the eye, the orange and blue filled circles connected with black lines show the binned median 
for the observation and simulations, respectively. In top left panel, the green filled circles connected with black lines show the model variation for 
Au06 as discussed in Section~\ref{catalogue}. The corresponding dots are not shown for the sake of clarity. The age bins were chosen in such a way that they 
contain approximately same numbers of stars.}
\label{fig8}
\end{figure*}

\subsection{Constraining radial stellar migration in the solar cylinder}
\label{migration}
Stars move radially in and out in the Galactic disc due to angular momentum transfer (`churning'), for instance from the bar and spiral arms, 
as well as due to scattering (`blurring'), for example from giant molecular clouds \citep[see][]{sell02,rosk08,scho09a,minc10,minc13,gran15}. 
The radial diffusion of stars caused by churning is known as radial stellar migration, whereas the one caused by blurring is termed kinematic or 
radial heating. 

There have been attempts to measure the strength of radial migration, as well as to quantify its importance in shaping the Galactic disc 
in comparison to kinematic heating. For instance, \citet{sand15} used action- and metallicity-based analytic distribution functions with a 
prescription for radial migration to fit the solar-neighbourhood chemodynamical data \citep[Geneva-Copenhagen Survey;][]{nord04} and the stellar 
density data \citep{gilm83}. They found the dispersion of angular momentum, $\sigma_L$, for 12 Gyr old stellar populations to be 1150 kpc km s$^{-1}$ 
(which can be translated to 939 kpc km s$^{-1}$ for 8 Gyr old populations using the relation $\sigma_L = 1150 \sqrt{\tau/12}$, where $\tau$ is 
the age of the population). Assuming a solar circular velocity of 235 km s$^{-1}$, this gives a radial migration of about 4.0 kpc for 8 Gyr old 
populations. Using the basic ideas from \citet{sand15}, \citet{fran18} developed a simple model by parametrizing the relevant physical processes 
including radial migration, and fitted it to the APOGEE DR12 \citep{alam15} low-$\alpha$ red clump stars. They found global radial migration of 
$3.6\pm0.1$ kpc for 8 Gyr old populations. By generalizing the model of \citet{fran18} in the 
direction of \citet{sand15} and using the data from APOGEE DR14, \citet{fran20} investigated the relative contributions of churning and blurring, 
and demonstrated that the impact of churning dominates. Furthermore, they found lower dispersion of angular momentum, 619 kpc km s$^{-1}$ for 
8 Gyr old populations, which corresponds to a lower value of radial migration of about 3.0 kpc compared to both \citet{sand15} and \citet{fran18}. 

\begin{figure*}
\includegraphics[scale=1]{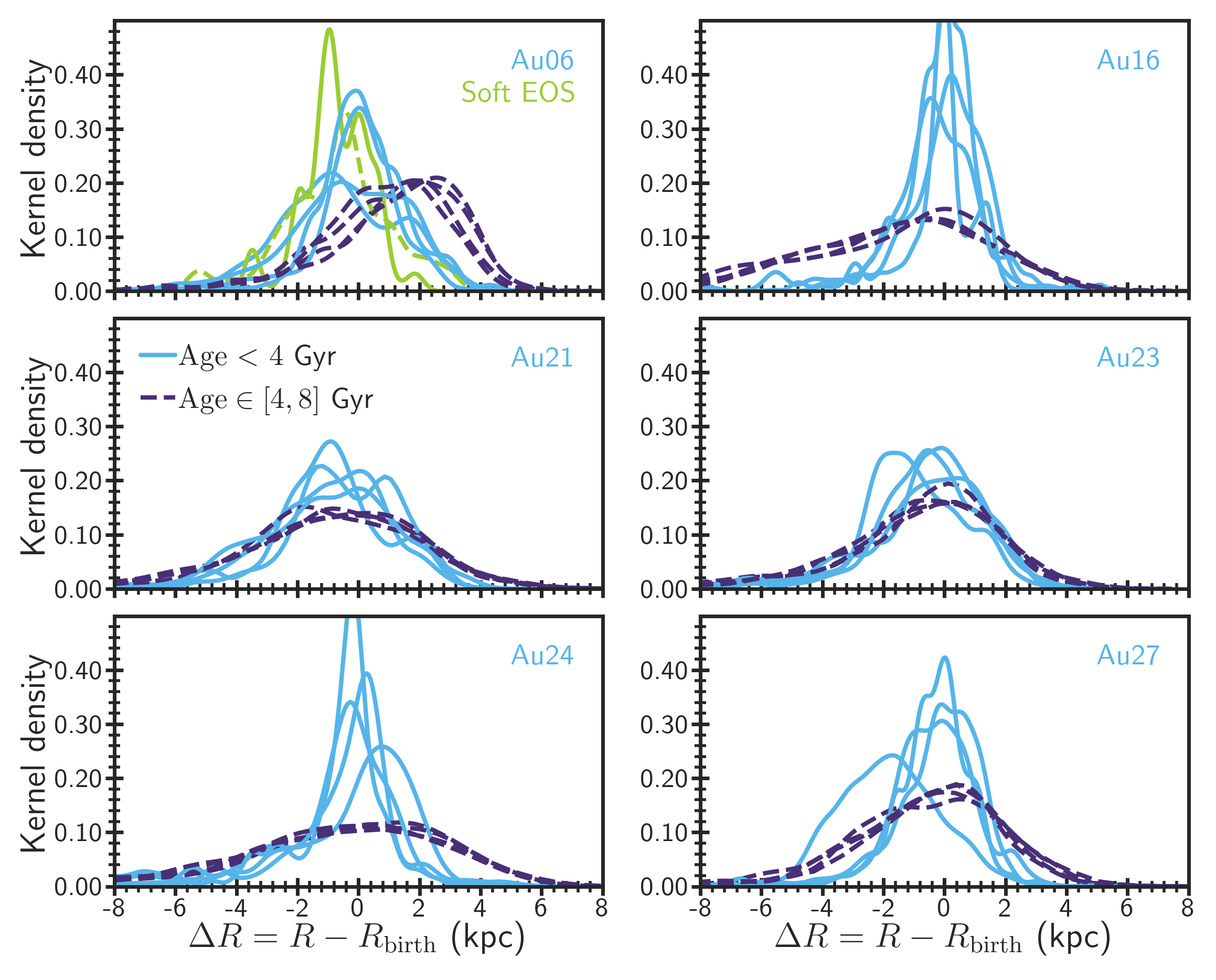}
\caption{Distributions of radial migration predicted by all of the six simulations. The different panels correspond to the six different 
simulations (see the legends). In each panel, the four continuous curves show the distributions for the young populations ($< 4$ Gyr) in ICC 
catalogues generated assuming different solar azimuths, while the four dashed curves demonstrate distributions for the old populations 
($\in [4, 8]$ Gyr). In top left panel, the green distributions show the model variation for Au06 as discussed in Section~\ref{catalogue}.}
\label{fig9}
\end{figure*}

The results of the above studies are crucial, and are being used to interpret the observations. For instance, \citet{shar20} used the dispersion of 
angular momentum from \citet{sand15} to explain the abundance dichotomy in $[{\rm Fe}/{\rm H}]-[\alpha/{\rm Fe}]$ plane. They also demonstrated 
the need for churning in reproducing the dichotomy in their study. However, since the above determinations of strength of radial migration rely 
heavily on numerous simplifying assumptions, and models have several limitations and caveats \citep[see Sections 3.1, 5.5 and 6.6 of][]{fran20}, 
it demands for constraints on radial migration from independent methods. Assuming our simulation sample is representative of Milky Way-like 
galaxies, we can use it to constrain radial migration in the Galaxy. In this section, we shall carefully analyse all six simulations to put an 
upper limit on the radial migration that took place for stars presently in the solar cylinder. 

\citet{sell02} defined radial diffusion caused by churning as radial migration. In practice, it is difficult to disentangle the contributions of 
churning and blurring, hence we defined radial migration of a star simply as the difference between its radial coordinates at present and birth 
locations, $R - R_{\rm birth}$. We emphasize that this definition includes contributions from both churning and blurring. Moreover, it has an 
ambiguity of measuring artificial radial migration for the stars born in highly eccentric orbits, i.e. this definition includes additional contributions
on top of those from churning and blurring.  

To alleviate the above ambiguity in the definition of radial migration, we shall consider only those stars in simulations that have relatively small 
orbital eccentricities, $e \lesssim 0.3$. In Figure~\ref{fig8}, we show eccentricities of the observed and simulation stars as a function of age. The 
eccentricities were estimated using {\tt galpy}\footnote{http://github.com/jobovy/galpy} assuming the Milky Way potential, {\tt MWPotential2014}, from 
\citet{bovy15}. Note that there are small differences between the {\tt MWPotential2014} and the simulation potentials, making the estimated
eccentricities prone to systematic uncertainties. However, these estimates serve the purpose for us as they are used only to separate the late 
`secular' evolution from the early turbulent evolution. As can be seen in the figure, older stars typically have large eccentricities, which is 
likely a result of their birth on highly eccentric orbits during the early merger dominated phase of galaxy formation and evolution. However, 
stars with ages less than 8 Gyr have median orbital eccentricities less than (or close to) 0.3 for all of the simulations, except possibly for Au21, 
for which stars of ages close to 8 Gyr have slightly larger eccentricities. Au21 underwent its most significant gas-rich merger at a lookback time 
of about 8 Gyr, leading to higher eccentricities around that time, while for the rest it happened at an earlier time 
\cite[see Figure~2 of][]{gran18a}. Therefore, we shall only consider observed and simulation stars with ages less than 8 Gyr in this section. 
Note that the spatial distribution and spectral type of the observed and simulation stars, as discussed in Section~\ref{selection}, remain 
similar.

\begin{figure*}
\includegraphics[scale=1]{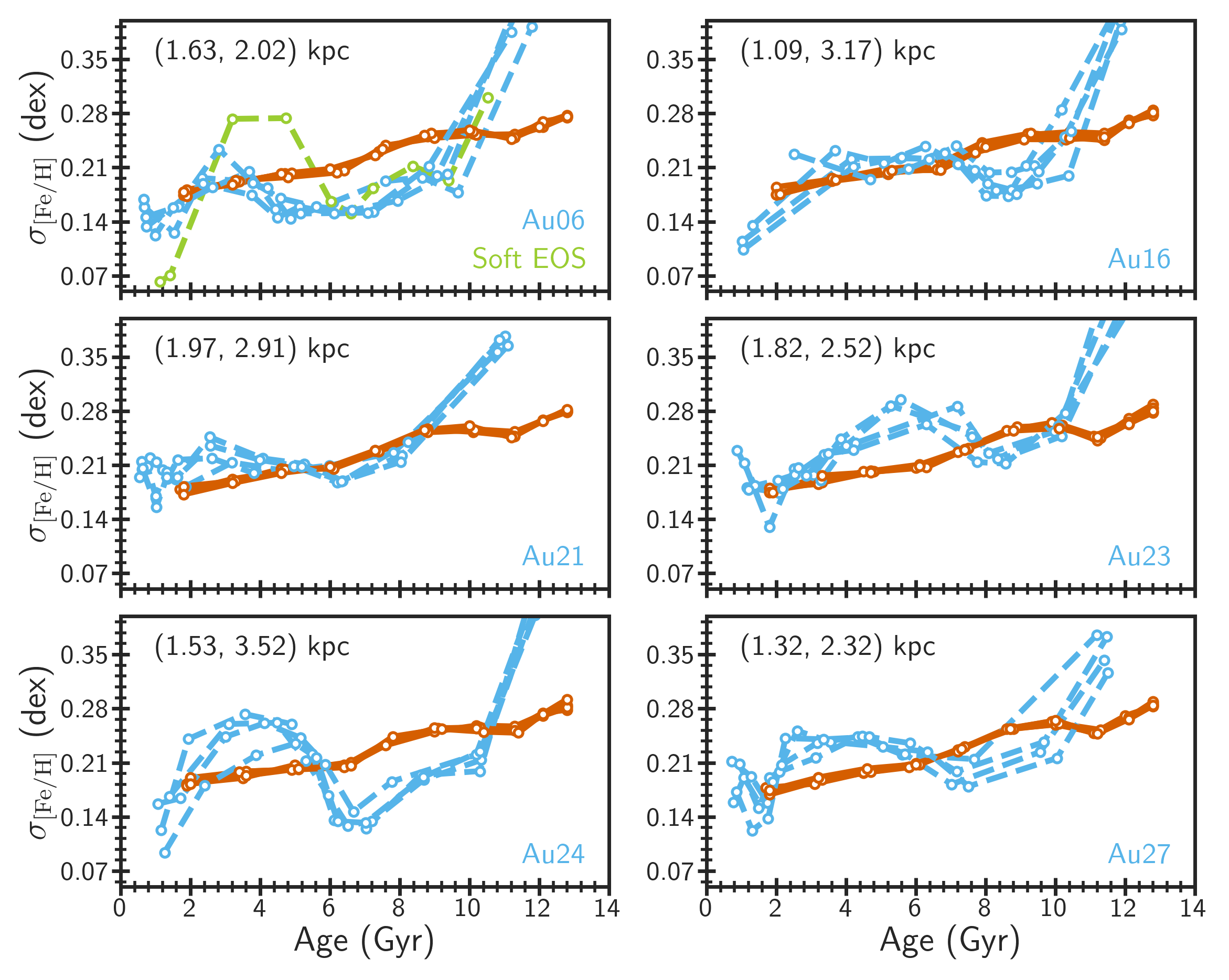}
\caption{Metallicity dispersion as a function of age. The different panels correspond to the six different simulations (see the legends in the bottom 
right corners). In each panel, the solid orange curves represent observed samples (see the corresponding text for details), while the other blue 
dashed curves correspond to the ICC catalogues generated assuming different solar azimuths. In top left panel, the green curve shows the model 
variation for Au06 as discussed in Section~\ref{catalogue}. The age bins were chosen in such a way that they contain approximately same numbers of stars. 
The legends in the top left corner list radial migrations (averaged over catalogues with different solar azimuths) predicted by the simulations for 
the young ($< 4$ Gyr) and old ($\in [4, 8]$ Gyr) stellar populations, respectively.} 
\label{fig10}
\end{figure*}

In Figure~\ref{fig9}, we show distributions of stellar migration predicted by all of the six simulations. To study its time dependence, we divided 
stars into two age bins: (1) young population with stellar ages less than 4 Gyr as shown by the continuous curves, and (2) old population 
with ages in the range [4, 8] Gyr as shown by the dashed curves. Note that, for all the simulations, the distributions are systematically wider 
for the older population than the younger one. This is expected because older stars have more time to migrate. In all panels, it is also 
interesting to note that the distributions corresponding to the four catalogues generated assuming different solar azimuths appear to be significantly 
different. This is mainly because of the differences in the distributions of the birth radius for the four catalogues (the distribution of the 
present-day radius remains approximately the same across catalogues because of the selection function). Note that since we have stars from different 
regions of the simulated galaxy for the four different catalogues, their birth profiles can differ due to the non-axisymmetric potential.

To quantify stellar migration, we again use the biweight midvariance statistic to estimate the standard deviation of the distributions in 
Figure~\ref{fig9}, and denote it with $\Delta{R}$. These estimates for both the young ($\Delta{R}_{\textless{4}}$) and old ($\Delta{R}_{\in [4,8]}$) 
populations are listed in Table~\ref{tab1} for all of the mock catalogues analysed in this study. The migrations predicted by the model variation
for Au06 with softer equation of state are 1.07 and 1.72 kpc for the young and old populations, respectively. These are relatively smaller than the 
corresponding values predicted by the fiducial hotter Au06 simulation (see Table~\ref{tab1}). We find maximum radial migration of 2.21 kpc (Au21)
and 3.70 kpc (Au24) for the young and old populations, respectively. We emphasize that these estimates include contributions from churning, blurring 
and from the fact that stars could have been born with non-zero eccentricity. Moreover, since all of the Auriga simulations have hotter kinematics
compared to the observations (see Figures~\ref{fig6} and \ref{fig7}), these values represent absolute upper limits on migration assuming its 
definition as given by \citet{sell02}, i.e. radial diffusion caused solely by churning.  

We wish to point out one caveat here: the particle positions and velocities at birth were not directly output from the simulations, but instead were 
calculated during post-processing from snapshots of cadence of about 250 Myr. Therefore, the computed birth positions and velocities have associated 
uncertainties, especially for stars born during mergers when substantial movements happen at shorter time-scales. However, since we consider only 
secular evolution in this section, such uncertainties should have only minor impact on the results. 

The signatures of radial migration have been also identified in stellar elemental abundances. Particularly, the metallicity dispersion in the 
age-metallicity distribution traces the strength of radial migration -- the larger the dispersion, the larger the migration 
\citep[see e.g.][]{hayw06,hayw08,scho09a}. In Figure~\ref{fig10}, we show the metallicity dispersion as a function of age for all of the simulations
and the observation. Recall that the observed orange curves may look slightly different depending on the catalogue and the simulation due to the 
selection function. To clearly see the relationship between the metallicity dispersion and radial migration, we note that the metallicity dispersion 
of Au16 is on average smaller than the other simulations as well as the observation for ages below 4 Gyr. This is in-line with our
expectation as this simulation predicts lowest value of radial migration for the young population. As we can see in the figure, the Au21 metallicity
dispersion on average matches closest with the observation for ages below 8 Gyr compared to the rest of the simulations. Therefore, it provides us 
estimates of radial migration of about 1.97 and 2.91 kpc for the young and old stellar populations, respectively. Note that these values are obtained
by averaging the radial migrations listed in Table~\ref{tab1} for Au21 over catalogues generated assuming different solar azimuths. We wish to
point out that the correlation between radial migration and metallicity dispersion depends on the underlying ISM metallicity gradient as a 
function of radius and time. For instance, a flatter radial gradient would cause a smaller metallicity dispersion even if there were a lot of 
migration. Although the absolute values of abundances predicted by the Auriga simulations differ significantly from the Milky Way, the gas phase 
radial metallicity gradient tends to be negative at most epochs. Therefore, the relative trends should be qualitatively consistent with expectations 
for the Milky Way (based on the inside-out formation scenario and the current explanations for the flat age-metallicity relation for example).

The above findings indicate that the inferred radial migrations in \citet{sand15,fran18} are too high, whereas the one found in \citet{fran20} 
is consistent with our study. Therefore, it would be interesting to see if a study similar to \citet{shar20} but with the spread in angular momentum 
from \citet{fran20} can still reproduce the abundance dichotomy.

\section{Summary and Conclusions}
We compared six high-resolution cosmological zoom-in simulations of the formation of Milky Way-mass galaxies from the Auriga project with a variety 
of observations from the APOGEE survey, the {\it Kepler} satellite and the {\it Gaia} mission. These simulations were a subset of a suite of 30 
simulations for which mock {\it Gaia} DR2 stellar catalogues were available. Our observed sample consisted of 7186 stars for which high-fidelity
spectroscopic, asteroseismic and astrometric data were all available. In order to make meaningful comparisons, we applied detailed selection function,
ensuring similar distributions of $J - K$ colour, $K$ magnitude and parallax for the observed and simulation stars. The process of applying the
selection function resulted in two sub-samples -- one of the observed sample, and the other of the mock catalogue at hand for comparison -- in which
stars had similar spatial distribution and spectral type. 

We compared the elemental abundances in $[{\rm Fe}/{\rm H}]-[{\rm Mg}/{\rm Fe}]$ plane. We found that simulations predict a variety of distributions
in this plane, some of which also show a dichotomy. The abundance dichotomy for Au27 looks particularly similar to the observed data. Although
simulations predict a dichotomy, we observe discrepancies when we look into the details of the observed and simulation abundance distributions. We 
also found that simulations tend to predict systematically higher metallicity compared to the observation. Some of these issues are likely to be 
related to the adopted yields in the simulations.

The simulation predictions for all three components of the Galactocentric velocity distributions were compared with the corresponding observed 
distributions. For all of the simulations, the predicted peak values of velocity distributions are in reasonable agreement with the observations. 
However, the fiducial Auriga simulations tend to systematically predict longer tails compared to the observed ones. The comparison of the vertical 
velocity dispersion as a function of age suggests significantly hotter kinematics in the simulated stars. However, a physics model variation that 
includes a softer equation of state for star forming gas produces a set of chemo-dynamical properties in much better agreement with observations, 
illustrating the power of this framework to constrain galaxy formation models in the face of high-dimensional Galactic data.

We used all six simulations together with the observed data to put constraints on radial stellar migration in the solar cylinder. We defined stellar 
migration as the difference between the radial coordinates at present and birth positions. Assuming that simulations in our sample are representative 
of Milky Way-like galaxies, we estimated upper limits of 2.21 and 3.70 kpc for the strength of radial migration for young ($< 4$ Gyr) and old 
($\in [4, 8]$ Gyr) stellar populations, respectively. The comparison of the metallicity dispersion as a function of age between the 
observation and simulations suggests the strength of radial migration to be about 1.97 and 2.91 kpc for the young and old populations, respectively. 
We emphasize that our definition of radial migration includes contributions from churning, blurring and from the fact that some stars may be born 
on elliptical orbits. This means migration caused solely by churning must be smaller than the above values. Since some of the simulations reproduce
the observed metallicity dispersion trend better in Figure~\ref{fig10} than the others, it can be a promising diagnostic for distinguishing the
formation and evolution histories.

Overall, we have developed a novel framework in which we address the challenging task of making systematic and unbiased comparisons between 
the rich contemporary observations and cosmological zoom-in simulations. We have also demonstrated the power of our technique by putting constraints on 
radial stellar migration. The current observed sample is limited to a small field-of-view (see Figure~\ref{fig1}), however this can be greatly 
expanded using the K2 and TESS asteroseismic data. Our framework has great potential, and we shall apply it to future generations of simulations with 
different galaxy formation models to constrain the formation of the Milky Way.

\section*{Acknowledgements}
We thank the anonymous referee for helpful comments. The authors would like to thank Volker Springel for inspiring discussions that led to the 
development of this project. Funding for the Stellar Astrophysics Centre is provided by The Danish National Research Foundation 
(Grant agreement no.: DNRF106). VSA acknowledges support from the Independent Research Fund Denmark (Research grant 7027-00096B) and the Carlsberg 
foundation (grant agreement CF19-0649).

\section*{Data Availability}
The data underlying this article will be shared on reasonable request to the corresponding author.


\appendix
\section{Results for the HITS catalogues and the ICC Catalogues with non-standard solar azimuths}

\begin{figure*}
\includegraphics[scale=0.89]{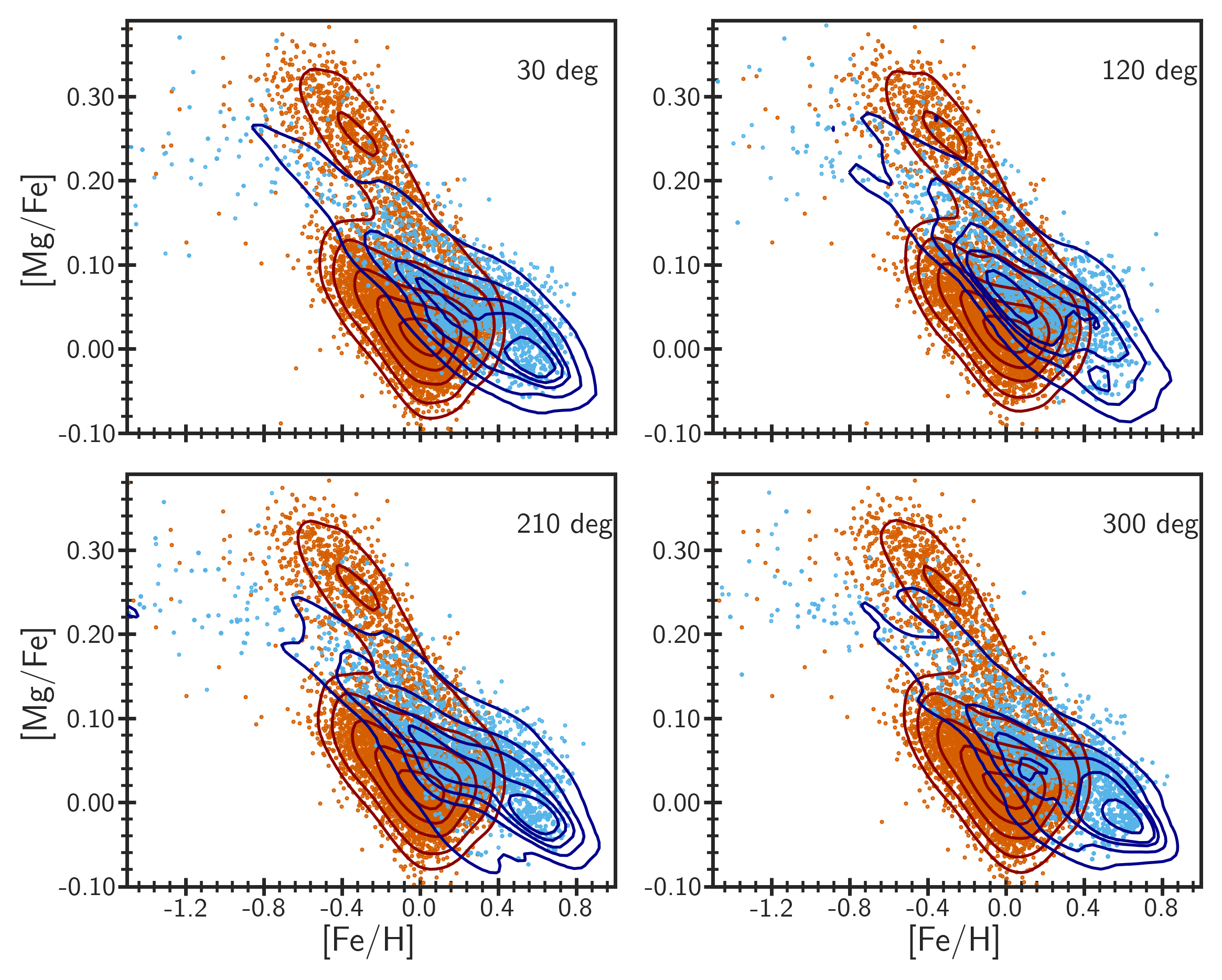}
\caption{Same as Figure~\ref{fig5} but contains only simulation Au06 with HITS catalogues generated assuming different solar azimuths (see the 
legends).}
\label{afig1}
\end{figure*}

\begin{figure*}
\includegraphics[scale=0.89]{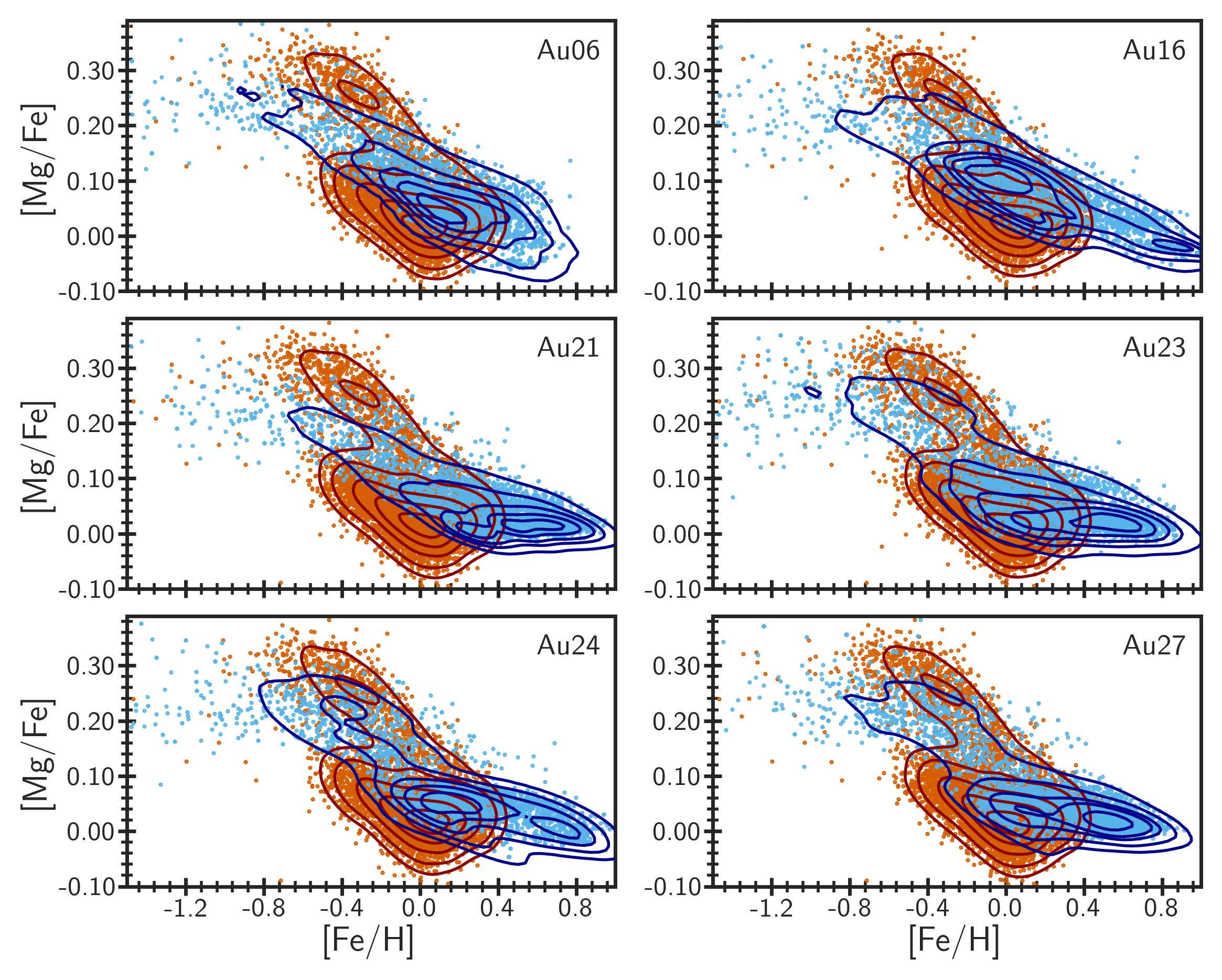}
\caption{Same as Figure~\ref{fig5} but with ICC catalogues generated assuming the solar azimuth at 120 deg behind the major axis of the bar.}
\label{afig2}
\end{figure*}

\begin{figure*}
\includegraphics[scale=0.89]{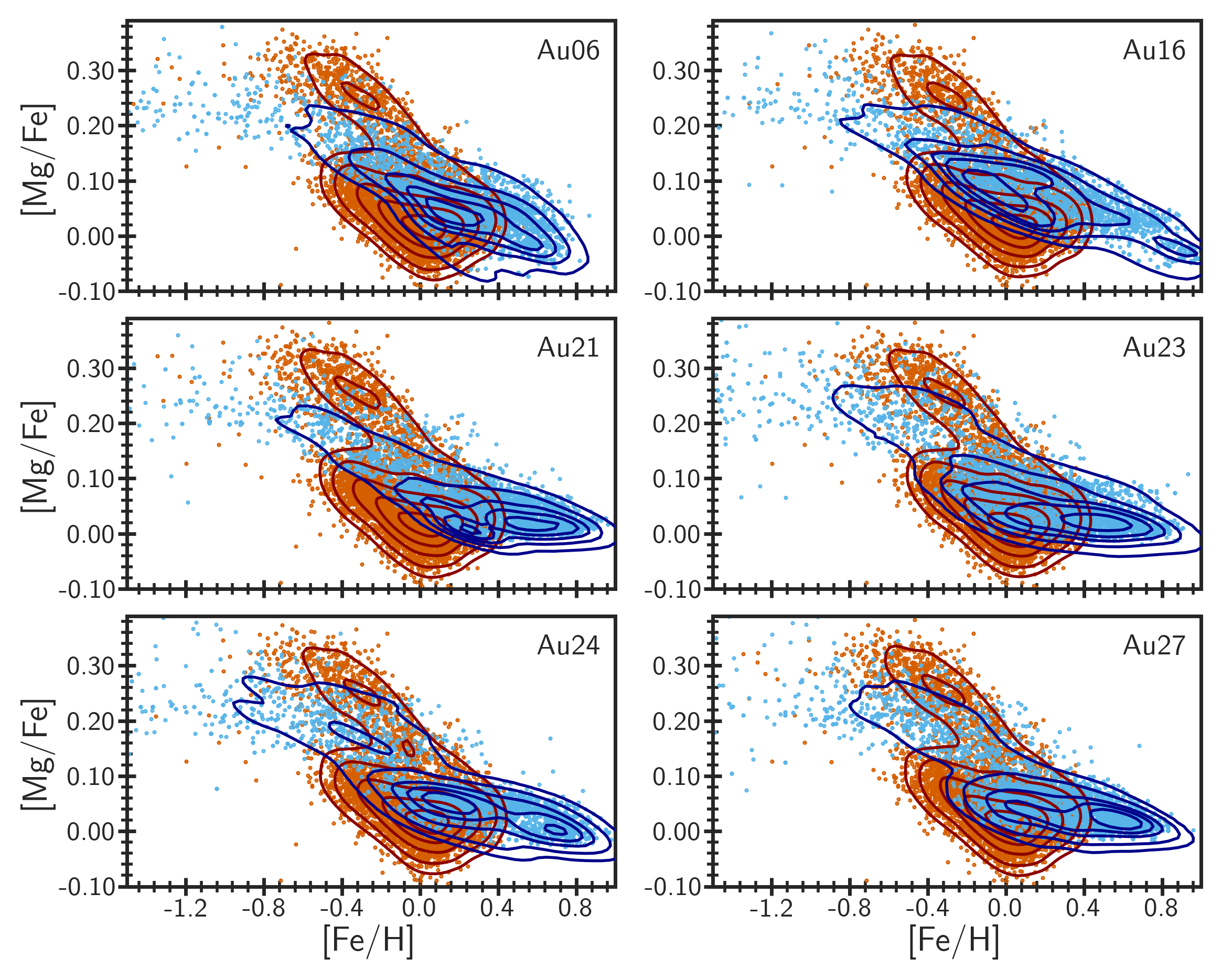}
\caption{Same as Figure~\ref{fig5} but with ICC catalogues generated assuming the solar azimuth at 210 deg behind the major axis of the bar.}
\label{afig3}
\end{figure*}

\begin{figure*}
\includegraphics[scale=0.89]{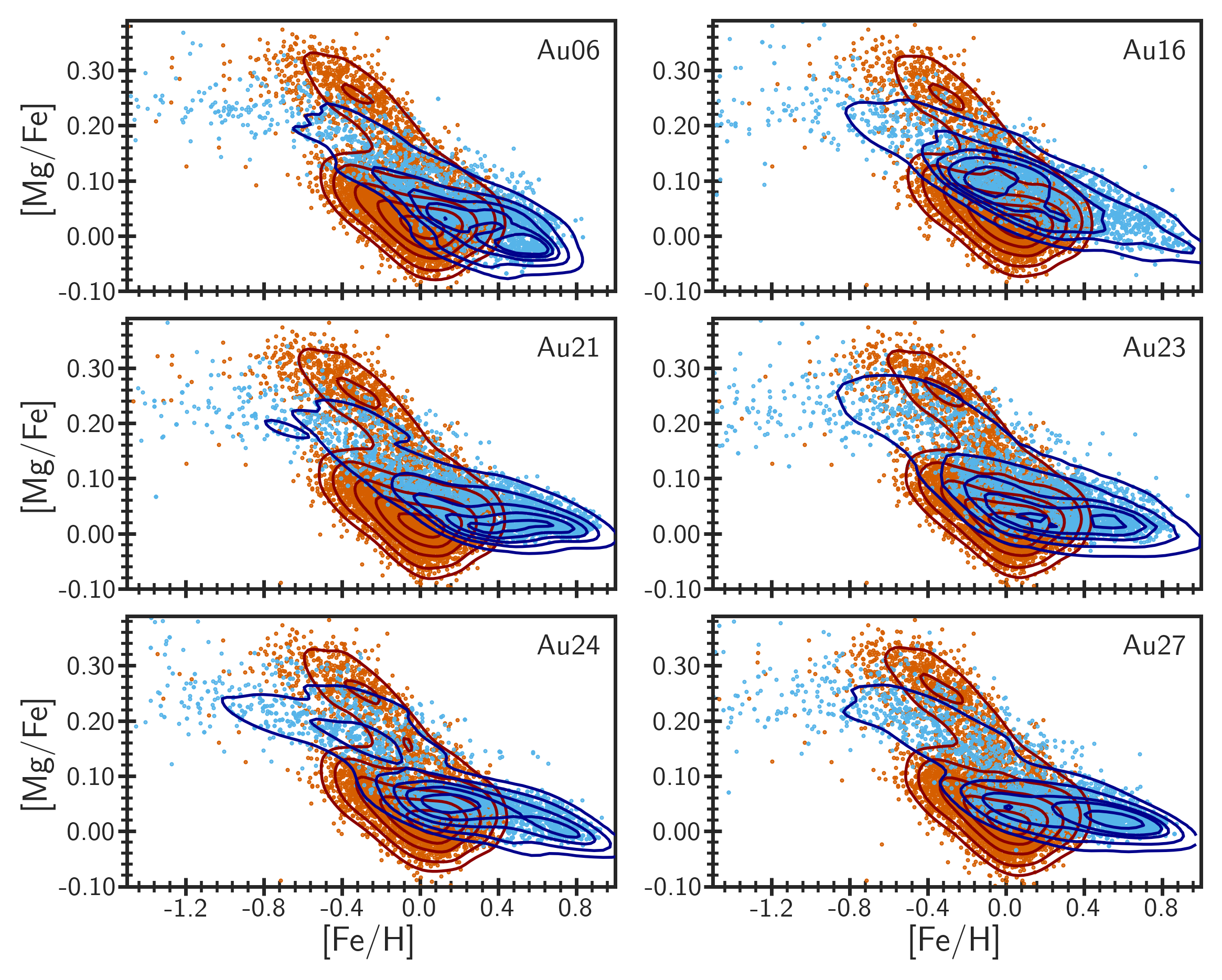}
\caption{Same as Figure~\ref{fig5} but with ICC catalogues generated assuming the solar azimuth at 300 deg behind the major axis of the bar.}
\label{afig4}
\end{figure*}

\begin{figure*}
\includegraphics[scale=0.90]{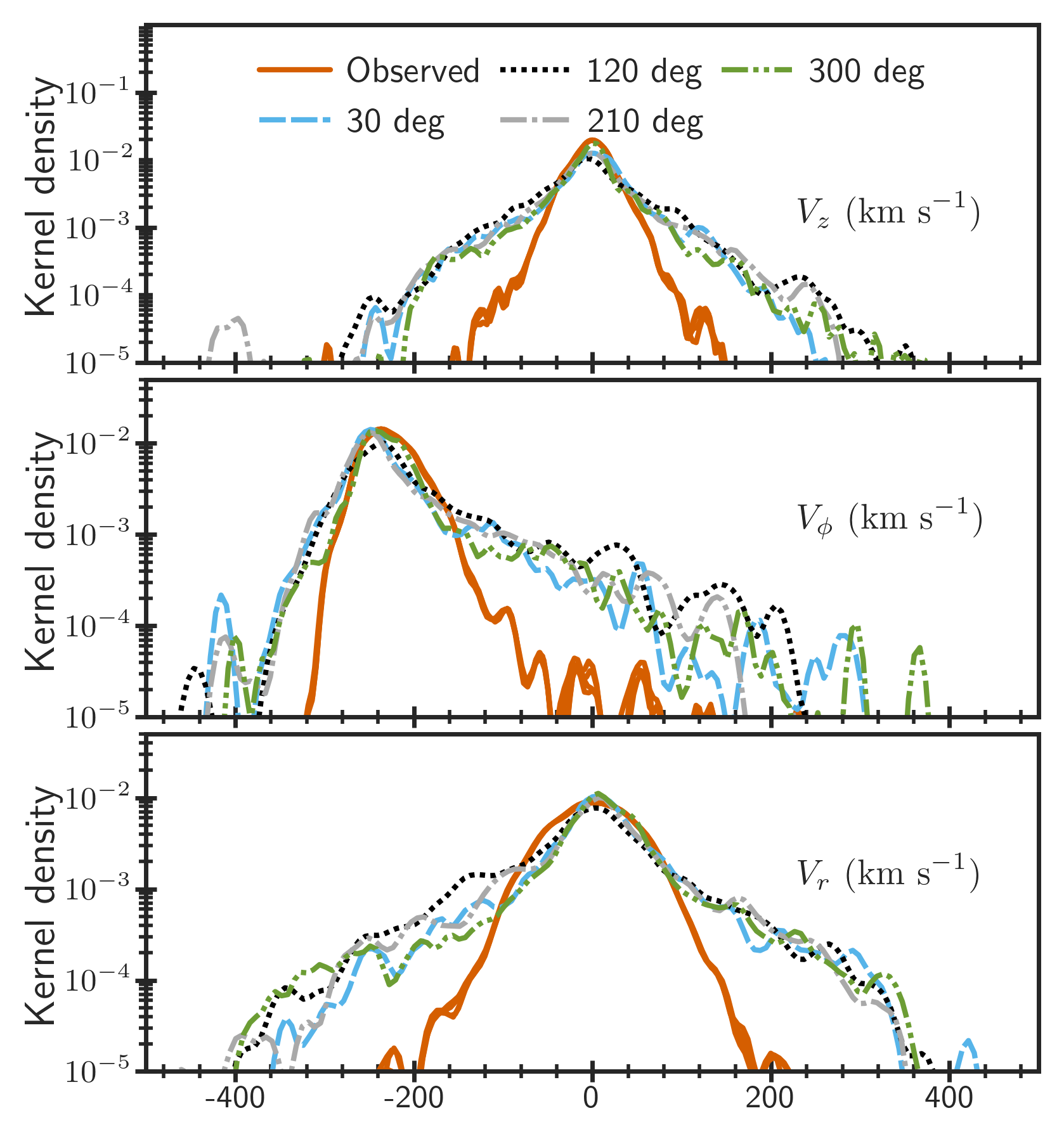}
\caption{Same as Figure~\ref{fig6} but contains only simulation Au06 with HITS catalogues generated assuming different solar azimuths (see the 
legends).}
\label{afig5}
\end{figure*}

\begin{figure*}
\includegraphics[scale=0.90]{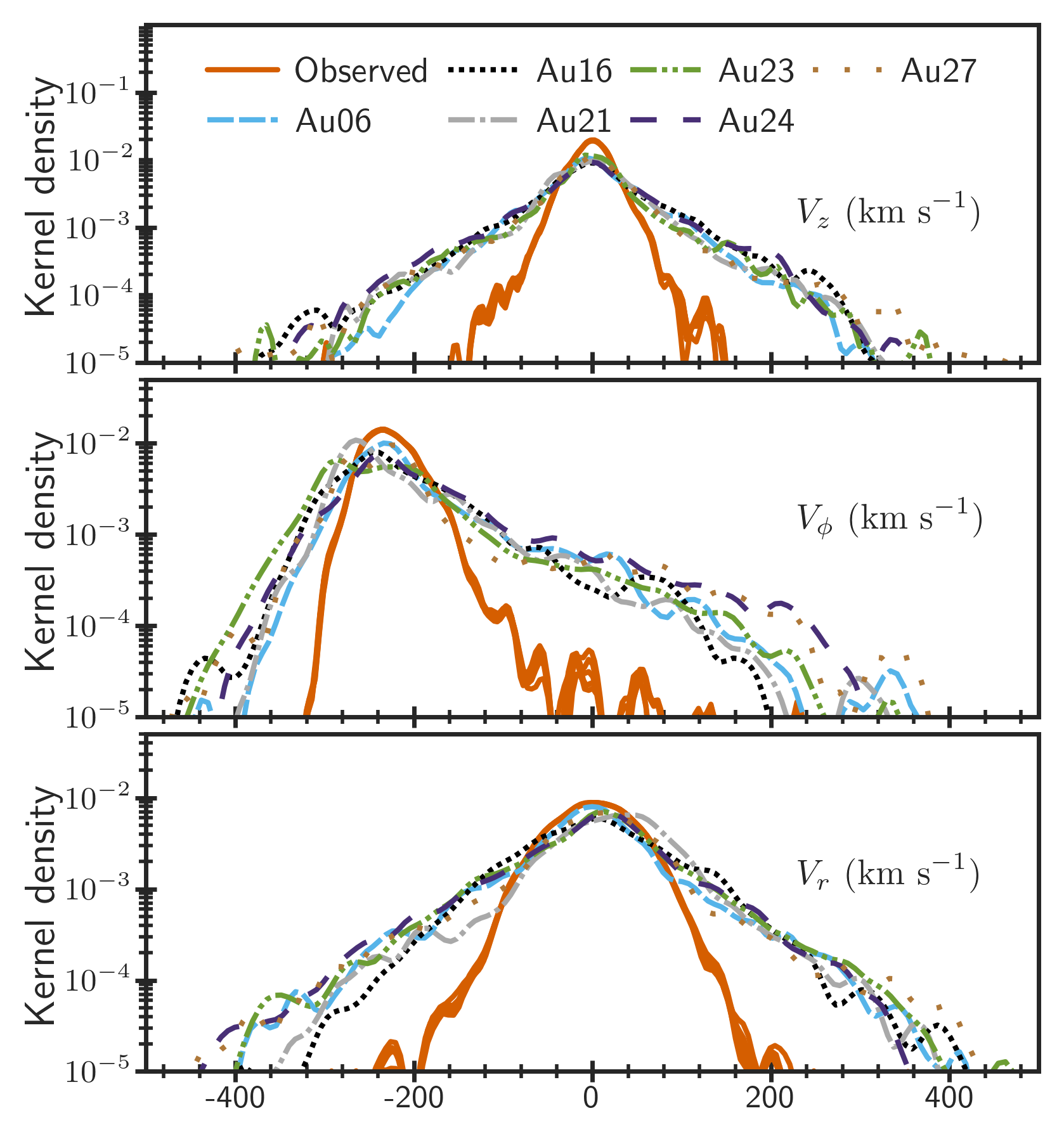}
\caption{Same as Figure~\ref{fig6} but with ICC catalogues generated assuming the solar azimuth at 120 deg behind the major axis of the bar.}
\label{afig6}
\end{figure*}

\begin{figure*}
\includegraphics[scale=0.90]{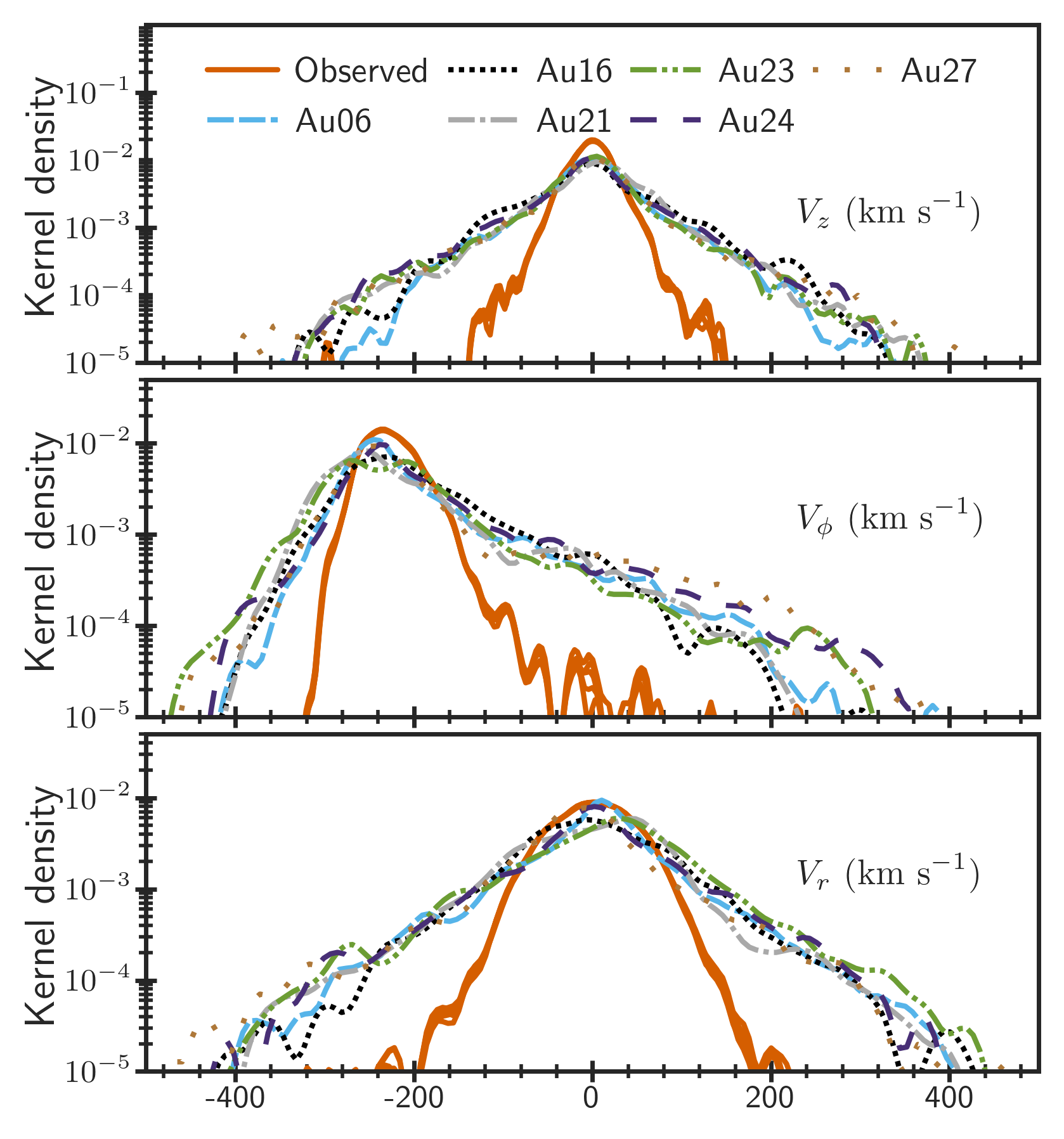}
\caption{Same as Figure~\ref{fig6} but with ICC catalogues generated assuming the solar azimuth at 210 deg behind the major axis of the bar.}
\label{afig7}
\end{figure*}

\begin{figure*}
\includegraphics[scale=0.90]{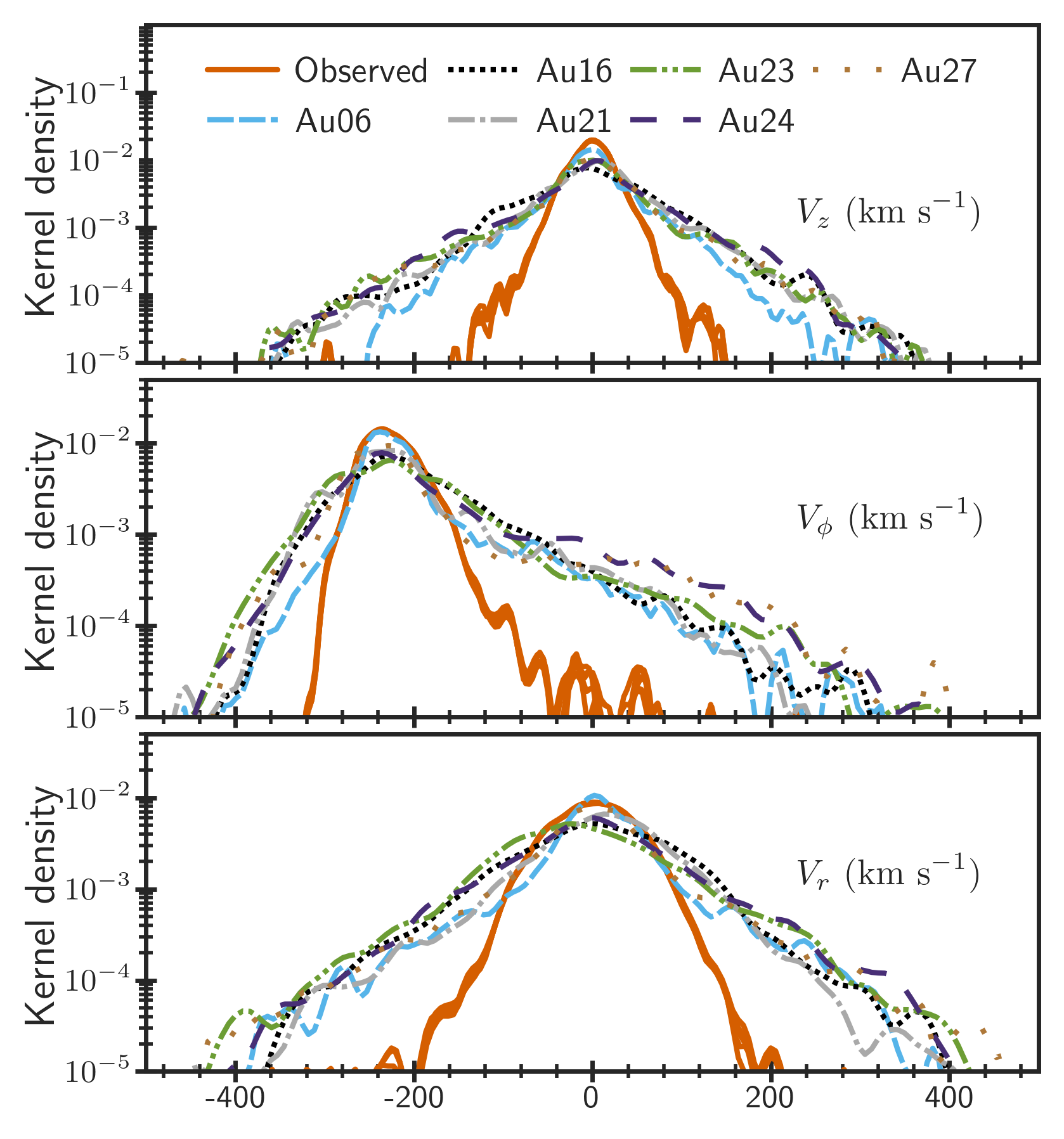}
\caption{Same as Figure~\ref{fig6} but with ICC catalogues generated assuming the solar azimuth at 300 deg behind the major axis of the bar.}
\label{afig8}
\end{figure*}

\bsp	
\label{lastpage}
\end{document}